\documentclass{aa}  

\usepackage{graphicx}
\usepackage{txfonts}
\usepackage{soul}
\usepackage{hyperref}
\hypersetup{
    colorlinks=true,
    citecolor=blue,
    urlcolor=blue
    }
    
\usepackage{multirow}
\usepackage{threeparttable, tablefootnote}

\graphicspath{{images/}}

\newcommand{\Msun}{\rm M_{\odot}}
\newcommand{\MsunYr}{\rm M_{\odot}~yr^{-1}}
\newcommand{\kms}{\rm km~s^{-1}}
\newcommand{\mum}{\mu \rm m}
\newcommand{\mm}{\rm mm}

\begin{document}

   \title{Modeling submillimeter galaxies in cosmological simulations: Contribution to the cosmic star formation density and predictions for future surveys}

   \titlerunning{Modeling submillimeter galaxies in cosmological simulations} 
   \authorrunning{Kumar et al. 2024}

   \author{Ankit Kumar\inst{1}\fnmsep\thanks{ankit4physics@gmail.com}
          \and
          M. Celeste Artale\inst{1}\fnmsep\thanks{maria.artale@unab.cl}
          \and
          Antonio D. Montero-Dorta\inst{2}
          \and
          Lucia Guaita\inst{1}
          \and
          Kyoung-Soo Lee\inst{3}
          \and
          Alexandra Pope\inst{4}
          \and
          Joop Schaye\inst{5}
          \and
          Matthieu Schaller\inst{5,6}
          \and
          Eric Gawiser\inst{7}
          \and
          Ho Seong Hwang\inst{8,9,10}
          \and
          Woong-Seob Jeong\inst{11}
          \and
          Jaehyun Lee\inst{11}
          \and
          Nelson Padilla\inst{13}
          \and
          Changbom Park\inst{12}
          \and
          Vandana Ramakrishnan\inst{3}
          \and
          Akriti Singh\inst{1}
          \and
          Yujin Yang\inst{11}
          }

   \institute{Universidad Andres Bello, Facultad de Ciencias Exactas, Departamento de Fisica y Astronomia, Instituto de Astrofisica, Fernandez Concha 700, Las Condes, Santiago RM, Chile
         \and
         Departamento de Física, Universidad Técnica Federico Santa María, Avenida Vicuña Mackenna 3939, San Joaquín, Santiago, Chile
         \and
         Department of Physics and Astronomy, Purdue University, 525 Northwestern Avenue, West Lafayette, IN 47907, USA
         \and
         Department of Astronomy, University of Massachusetts, Amherst, MA 01003, USA
         \and
         Leiden Observatory, Leiden University, PO Box 9513, 2300 RA Leiden, the Netherlands
         \and
         Lorentz Institute for Theoretical Physics, Leiden University, PO box 9506, 2300 RA Leiden, the Netherlands
         \and
         Physics and Astronomy Department, Rutgers, The State University, Piscataway, NJ 08854, USA
         \and
         Department of Physics and Astronomy, Seoul National University, 1 Gwanak-ro, Gwanak-gu, Seoul 08826, Republic of Korea
         \and
         SNU Astronomy Research Center, Seoul National University, 1 Gwanak-ro, Gwanak-gu, Seoul 08826, Republic of Korea
         \and
         Australian Astronomical Optics - Macquarie University, 105 Delhi Road, North Ryde, NSW 2113, Australia
         \and
         Korea Astronomy and Space Science Institute, 776 Daedeokdae-ro, Yuseong-gu, Daejeon 34055, Republic of Korea
         \and
         Korea Institute for Advanced Study, 85 Hoegi-ro, Dongdaemun-gu, Seoul 02455, Republic of Korea
         \and
         Instituto de Astronomía Teórica y Experimental (IATE), CONICET-UNC, Laprida 854, X500BGR, Córdoba, Argentina
             }


 
  \abstract
   {Submillimeter galaxies (SMGs) constitute a key population of bright star-forming galaxies at high redshift. These galaxies  challenge galaxy formation models, particularly in reproducing their observed number counts and redshift distributions. 
   Furthermore, although SMGs contribute significantly to the cosmic star formation rate density (SFRD), their precise role remains uncertain.
   Upcoming surveys, such as the Ultra Deep Survey with the TolTEC camera, are expected to offer valuable insights into SMG properties and their broader impact in the Universe.
   }
   {Robust modeling of SMGs in a cosmologically representative volume is necessary to investigate their nature in preparation for next-generation submillimeter surveys. Here we test different parametric models for SMGs in large-volume hydrodynamical simulations, assess their contribution to the SFRD, and build expectations for future submillimeter surveys.}
   {We implement and test parametric relations derived from radiative transfer (RT) calculations across three cosmological simulation suites: EAGLE, IllustrisTNG, and FLAMINGO. Particular emphasis is placed on the FLAMINGO simulations due to their large volume and robust statistical sampling of SMGs. 
   Based on the model that best reproduces observational number counts, we forecast submillimeter fluxes within the simulations, analyze the properties of SMGs, and evaluate their evolution over cosmic time.
   }
   {Our results show that the FLAMINGO simulation reproduces the observed redshift distribution and source number counts of SMGs without requiring a top-heavy initial mass function. On the other hand, the EAGLE and IllustrisTNG simulations show a deficit of bright SMGs. 
   We find that SMGs with S$_{850} > 1$ mJy contribute up to $\sim 27\%$ of the cosmic SFRD at $z \sim 2.6$ in the FLAMINGO simulation, consistent with recent observations.
   Flux density functions reveal a rise in SMG abundance from $z = 6$ to $z = 2.5$, followed by a sharp decline in the number of brighter SMGs from $z = 2.5$ to $z = 0$. 
   Leveraging the SMG population in FLAMINGO, we forecast that the TolTEC UDS will detect $\sim$80000 sources over $0.8~{\rm deg}^2$ at $1.1~\mm$ (at the $4\sigma$ detection limit), capturing about 50\% of the cosmic SFRD at $z \sim 2.5$.
  }
   {}

   \keywords{Galaxies: formation -- Galaxies: evolution -- Submillimeter: galaxies -- Infrared: galaxies -- Galaxies: high-redshift}

   \maketitle
%

\section{Introduction}
\label{sec:introduction}
Submillimeter galaxies (SMGs), also known as dusty star-forming galaxies (DSFGs), are a class of galaxies that predominantly emit in the submillimeter range of the electromagnetic spectrum. SMGs are typically observed at redshifts $z \simeq 2-3$, and are considered the high redshift counterparts of ultraluminous infrared galaxies (ULIRGs). They are among the most intensely star-forming galaxies in the Universe, with star formation rates (SFRs) often exceeding $100-1000$ solar masses per year \citep[e.g.,][]{Barger.etal.1998, Chapman.etal.2005, Barger.etal.2012, Hezaveh.Marrone.etal.2013}.
 
Their high SFR and dusty environment made SMGs an important probe to understand galaxy formation and evolution \citep[e.g.,][]{Smail.etal.1997, Barger.etal.1998}. 
Different reports suggest that the presence of abundant cold molecular gas and interactions/mergers between galaxies fuel intense star formation in SMGs \citep[e.g.,][]{Frayer.etal.1999, Ivison.etal.2000, Tacconi.etal.2008, Engel.etal.2010, Barger.etal.2012, Bothwell.etal.2013, Narayanan.etal.2015}.
Furthermore, the dense interstellar medium (ISM) found in SMGs, combined with dust presence, protects molecular gas from ionizing radiation, promoting efficient cooling and star formation \citep[][]{Casey.etal.2014}.
This is evidenced, for instance, by interferometric observations of CO in SMGs at $z \approx 2-4$ showing high molecular gas content ($\sim 5\times 10^{10}~\Msun$) within the central 2 kpc \citep{Bothwell.etal.2013}. The CO velocity gradient of $\approx 500-800~\kms$ suggests that SMGs are merging galaxies or rotating disks \citep{Greve.etal.2005, Tacconi.etal.2008}. This large amount of molecular gas fuels star formation in SMGs for $20-100$ Myr \citep{Yang.etal.2017, McAlpine.etal.2019}. 

SMGs account for a significant part of the cosmic star formation rate density (SFRD) at high redshift ($z=2-3$) and contribute to the cosmic infrared background \citep{Chapman.etal.2003, Chapman.etal.2005, Dunlop.etal.2017, Michalowski.etal.2017, Smith.etal.2017}. SMGs lie at the high mass end of the star forming main sequence or above it, with median SFR $\sim 300 ~\Msun/yr$ and median stellar mass $\sim 10^{11}~\Msun$ \citep{da_Cunha.etal.2015, Michalowski.etal.2017, Miettinen.etal.2017}. At the same star formation rate, the host halos of SMGs at high redshift are more massive than the host halos at low redshift \citep{Magliocchetti.etal.2013, Wilkinson.etal.2017}. The typically high mass halos of SMGs \citep[$\sim 10^{13}~\Msun$;][]{Chen.etal.2016, Lim.etal.2020} are thought be the progenitors of present day massive elliptical galaxies \citep{Simpson.etal.2014, Toft.etal.2014, Wilkinson.etal.2017, Gomez-Guijarro.2018, Valentino.etal.2020}.


A key feature that makes SMGs significant targets to study at high redshifts, is their nearly constant flux density as a function of redshift from $z=0.5$ to $z=10$ at submillimeter wavelengths, which allows the detection of more distant galaxies compared to optical bandwidths \citep{Blain.1997}.
Furthermore, the strong clustering of SMGs at $z=2-3$ suggests that they are good tracers of large and dense structures at high redshift \citep{Blain.etal.2004, Hickox.etal.2012, Wilkinson.etal.2017}, making them potential protocluster tracers \citep{Zhang.etal.2022}.

In the last three decades, significant progress has been made in understanding the nature of SMGs using observational surveys and theoretical models. From the observational side, surveys from single-dish telescopes contributing in this direction are SCUBA \citep{Smail.etal.1997, Hughes.etal.1998}, SCUBA-2 \citep{Chapman.etal.2005, Coppin.etal.2006, Geach.etal.2017}, Large APEX BOlometer CAmera \citep[LABOCA:][]{Siringo.etal.2009, Weiss.etal.2009}, and AzTEC \citep{Scott.etal.2008, Hatsukade.etal.2011}.
However, their detailed studies are still poorly understood due to difficulties in optical counterpart identification, spectroscopic redshift measurements, and modeling their observed characteristics (e.g., high SFRs, source number counts, redshift distributions, and multiplicity, among others). 
On the other hand, the high angular resolution ($\approx 1.5$ arcsec FWHM) of interferometric telescopes such as IRAM PdBI (Plateau de Bure Interferometer), SMA (Sub-Millimetre Array) and ALMA (Atacama Large Millimetre Array) has enabled more detailed studies of the SMG population. However, interferometric telescopes map small areas in the sky, which limits the capabilities for mapping a large statistical sample of the SMG population.  
In this context, the forthcoming TolTEC camera \citep{Wilson.etal.2020} at the 50-m Large Millimeter Telescope \citep[LMT: ][]{Hughes.etal.2020} will provide new insights into dust-obscured star-forming galaxies. 
With mapping speeds exceeding 2~deg$^2$/mJy$^2$/hr and offering high angular resolution (5 arcsec FWHM), it will enable the execution of two public legacy surveys at $1.1~\mm$, $1.4~\mm$, and $2.0~\mm$ wavelengths: the $\rm 0.8~deg^{2}$ Ultra Deep Survey (UDS) and the $\rm 40-60~deg^{2}$ Large Scale Survey (LSS).

Theoretical modeling of SMGs encompasses different methodologies. 
One approach uses dark matter-only simulations and semi-analytical models (SAMs) or semi-empirical relationships
\citep[e.g.,][]{Baugh.etal.2005, Somerville.etal.2012, Munoz_Arancibia.etal.2015, Safarzadeh.etal.2017, Lagos.etal.2020, Nava-Morena.etal.2024}. This approach has proven to be an exceptional resource due to its substantial datasets and suitability to study the large-scale distribution of SMGs.
On the other hand, hydrodynamical simulations have shown to be complementary to SAMs and semi-empirical models, as they can reproduce the internal structures of galaxies in a self-consistent way. For modeling SMGs, some authors post-process snapshots from hydrodynamical simulations with dust radiative transfer codes \citep[e.g.,][]{Lovell.etal.2021, Cochrane.etal.2023, McAlpine.etal.2019}, while others use an analytical formalism \citep[e.g.,][]{Dave.etal.2010,Shimizu.etal.2012}. In addition, high-resolution hydrodynamical simulations of isolated and/or interacting galaxies have been shown to be a great complement with the advantage of better resolving the ISM, but lack knowledge of the cosmological environment \citep[e.g.,][]{Chakrabarti.etal.2008, Narayanan.etal.2010, Hayward.etal.2011}. 
Another alternative is to use parametric models derived from 3D dust radiative transfer (RT) calculations on idealized high-resolution simulations \citep[e.g.,][]{Hayward.etal.2011, Hayward.etal.2013, Cochrane.etal.2023}. 
This method calculates the submillimeter flux density based on the star formation rate (SFR) and dust mass of each simulated galaxy and has several advantages. First, it significantly reduces computational costs compared to radiative transfer modeling. Furthermore, it can be implemented in large-volume cosmological hydrodynamical simulations where the mass and spatial resolution are too low to perform radiative transfer modeling. 

All of these approaches have faced challenges in modeling and reproducing the observed source number counts along with the redshift distribution of SMGs. 
Furthermore, the blending of multiple SMGs, particularly from single-dish observational data, causes an overprediction in the number counts  \citep[as shown by, e.g.,][]{Karim.etal.2013,Danielson.etal.2017,Stach.etal.2018}.
Upcoming SMG surveys, such as the UDS and LSS with the TolTEC camera, will yield new observational data crucial for interpreting the nature and origins of these galaxies. In this context, investigating SMG modeling and especially the significance of subgrid physics, is pertinent. 

In this work, we investigate the SMG population within three cosmological simulations: EAGLE, IllustrisTNG, and FLAMINGO, and generate predictions for the upcoming TolTEC surveys. Our study is framed within the context of the ODIN (One-hundred-deg$^2$ DECam Imaging in Narrowbands) survey, which maps large-scale structures and protocluster regions at $z = 2.4, 3.1,$ and 4.5 using Ly$\alpha$-emitting galaxies (LAEs; \citealt{Lee2024, Andrews2024}). This work focuses on the modeling of SMGs in cosmological simulations, complementing the ODIN analysis. The role of SMGs as protocluster tracers, in comparison to LAEs, will be tested in future studies (Kumar et al., in prep).

We begin by testing parametric models derived from radiative transfer calculations in the EAGLE cosmological simulation \citep{Schaye.etal.2015, Crain.etal.2015, Camps.etal.2018}, benchmarking these models against a sample of 892 observational SMGs. We then apply the parametric models to the three cosmological simulations, comparing their predictions with observed redshift distributions and source number counts. Leveraging the large cosmological volume of the FLAMINGO simulation \citep{Schaye.etal.2023, Kugel.etal.2023}, we estimate the contribution of SMGs to the cosmic star formation rate density, examine the evolution of their flux density function, and provide forecasts for upcoming TolTEC surveys \citep{Wilson.etal.2020}.

This paper is organized as follows. Section~\ref{sec:simulations} briefly describes the three state-of-the-art hydrodynamical simulations (EAGLE, IllustrisTNG, and FLAMINGO) used in this work. In Section~\ref{sec:smg_modeling}, we detail the modeling of the SMG population in cosmological simulations, and our tests on simulated and observed SMGs to choose a robust model. Section~\ref{sec:results} presents our results along with a discussion showing redshift distributions, source number counts, contribution to cosmic star formation rate density, and evolution of flux density functions. Section~\ref{sec:toltec_prediction} discusses our predictions for TolTEC surveys. Finally, Section~\ref{sec:conclusions} summarizes the main results of this work.

\section{Hydrodynamical simulations}
\label{sec:simulations}
In this work, we utilize three hydrodynamical cosmological simulations: EAGLE, IllustrisTNG, and FLAMINGO. The main focus of this paper is to use the cosmologically representative volume of the FLAMINGO simulation for the modeling of SMGs. This section briefly summarizes the three simulations.

\subsection{EAGLE}
\label{sec:eagle_sims}
EAGLE (Evolution and Assembly of GaLaxies and Their Environments) is a suite of cosmological hydrodynamical simulations from the Virgo Consortium\footnote{\url{https://virgo.dur.ac.uk/}}. All EAGLE simulations are executed using the \textsc{gadget3} code \citep{Springel.2005}, but with a new hydrodynamics solver and new subgrid prescriptions. EAGLE's subgrid physics includes element-by-element radiative cooling and photoheating for the most important 11 elements (H, He, C, N, O, Ne, Mg, Si, S, Ca, and Fe), metallicity-dependent star formation, stellar mass loss, supernovae, supermassive black holes, and AGN feedback. The EAGLE project and the detailed implementation of subgrid physics are described in \cite{Schaye.etal.2015} and \cite{Crain.etal.2015}. These simulations are calibrated to reproduce observed galaxy sizes, the galaxy stellar mass function, and the black hole mass $-$ stellar mass relation, all at $z=0$. The EAGLE simulations use the Chabrier initial mass function \citep{Chabrier.etal.2003}.

The main EAGLE simulations comprise three cosmological boxes that have 25, 50, and 100 comoving Mpc (cMpc) side lengths. The primary output of the simulations is stored in 29 snapshots starting from redshift $z=20$ to $z=0$. There are also 400 snapshots with a reduced number of properties (called snipshots) in the same redshift range. The halos (groups) and subhalos (galaxies) are respectively identified using the friends of friends (\textsc{FoF}) \citep{Press.Davis.1982, Davis.etal.1985} and \textsc{Subfind} \citep{Springel.etal.2001} algorithms. \textsc{FoF} is performed on dark matter particles only, while \textsc{Subfind} is performed on all the particles within halos to find gravitationally bound subhalos of particles.
The data is publicly available through SQL query on the Virgo Consortium database website\footnote{\url{https://virgodb.dur.ac.uk/}} \citep[for details, see][]{McAlpine.etal.2016}. Properties of galaxies, e.g., stellar mass, half-mass radius, star formation rate, and metallicity are computed using particles within a 3D aperture of radius 30 physical kpc (pkpc).

We will use the largest EAGLE box, which has a 100 cMpc side length, named RefL0100N1504 for validating our SMG population synthesis. It consists of $2 \times 1504^{3}$ particles with gas mass $1.81 \times 10^{6}~\Msun$ and dark matter mass $9.7 \times 10^{6}~\Msun$. Table~\ref{tab:sim_param_tab} shows the cosmological parameters for the EAGLE simulation and describes our selection cuts imposed on the sample galaxies.

\subsection{IllustrisTNG}
\label{sec:illustrisTNG_sims}
IllustrisTNG (in short, TNG) is a suite of magneto-hydrodynamical cosmological simulations. All TNG simulations are performed using the moving-mesh code \textsc{arepo} \citep{Springel.2010} and run from redshift $z=127$ to $z=0$. These simulations include metal cooling, star formation, stellar evolution, chemical evolution (of H, He, C, N, O, Ne, Mg, Si, Fe), stellar feedback, black hole seeding, AGN feedback, and magnetic fields. The TNG galaxy formation model is described in \cite{Weinberger.etal.2017} and \cite{Pillepich.etal.2018}. These simulations are calibrated to match the stellar mass function, the stellar mass $-$ stellar size relation, the stellar-to-halo mass relation, the black hole $-$ galaxy mass relation, mass $-$ metallicity relation, the gas fraction in massive groups all at $z=0$, and the cosmic star formation rate density for $z<10$. The IllustrisTNG simulations use the Chabrier initial mass function \citep{Chabrier.etal.2003}.

TNG includes three cosmological boxes of approximately 300, 100 and 50 cMpc side lengths referred to as TNG300, TNG100 and TNG50, respectively. Here, TNG300 is the lowest resolution box with the largest side length, suitable for studying galaxy clustering, whereas TNG50 is the highest resolution box with the smallest side length and is suitable for studying galaxy formation in detail. There are 100 snapshots stored for each simulation from $z=20$ to $z=0$. Similarly to the EAGLE, halos are searched using the \textsc{FoF} algorithm and subhalos are found using the \textsc{Subfind} algorithm. The merger trees are available from two algorithms, namely \textsc{SubLink} \citep{Rodriguez-Gomez.etal.2015} and \textsc{LHaloTree} \citep{Springel.etal.2005}. All data products are publicly available on the TNG website\footnote{\url{https://www.tng-project.org/}} \citep{Nelson.etal.2019}.

In this study, we utilise TNG100 and TNG300 galaxies. The total number of resolution elements in TNG100 is $2 \times 1820^{3}$, and in TNG300 it is $2 \times 2500^{3}$. TNG100 has dark matter and gas particle masses of, respectively, $7.5 \times 10^{6}~\Msun$ and $1.4 \times 10^{6}~\Msun$. On the other hand, TNG300 has dark matter and gas particle masses of, respectively, $5.9 \times 10^{7}~\Msun$ and $1.1 \times 10^{7}~\Msun$. Table~\ref{tab:sim_param_tab} lists the cosmological parameters for the IllustrisTNG simulations and summarises our selection cuts imposed on the sample galaxies.

\begin{table}
    \centering
    \caption{Cosmological parameters of different simulations and constraints on data used in this work.}
    \begin{threeparttable}
    \begin{tabular}{|c|c|c|c|}
    \hline
    parameter & EAGLE & IllustrisTNG & FLAMINGO\\
     &  &  & (L1\_m8)\\
    \hline
    \multicolumn{4}{|c|}{(cosmological parameters)} \\ 
    \hline
    $\Omega_{\Lambda, 0}$ & 0.693 & 0.6911 & 0.694\\
    $\Omega_{m, 0}$ & 0.307 & 0.3089 & 0.306\\
    $\Omega_{b, 0}$ & 0.04825 & 0.0486 & 0.0486\\
    $\sigma_{8}$ & 0.8288 & 0.8159 & 0.807\\
    $n_{s}$ & 0.9611 & 0.9667 & 0.967\\
    $h$ & 0.6777 & 0.6774 & 0.681\\
    \hline
    \multicolumn{4}{|c|}{(data used in this work)} \\ 
    \hline
    redshift$^{(1)}$ & $z \ge 0.5$ & $z \ge 0.5$ & $z \ge 0.5$ \\
    subhalo type$^{(2)}$ & central & central & central \\
    stellar mass$^{(3)}$ & $ \ge 10^{8.5}~\Msun$ & $ \ge 10^{8.5}~\Msun$ & $ \ge 10^{9.1}~\Msun$ \\
    aperture size$^{(4)}$ & 30~pkpc & 30~pkpc & 30~pkpc \\
    \hline
    \end{tabular}
    \end{threeparttable}
    {(1) Though all three simulations are available until $z=0$, we use only $z \ge 0.5$ snapshots because the contribution of SMGs at low-redshift is negligible. (2) Central subhalo is the most massive subhalo of any group, (3) Stellar mass of the subhalo, (4) Size of spherical aperture placed on subhalo center for measurements [in physical kpc].}
    \label{tab:sim_param_tab}
\end{table}

\subsection{FLAMINGO}
\label{sec:flamingo_sims}
FLAMINGO (Full-hydro Large-scale structure simulations with All-sky Mapping for the Interpretation of Next Generation Observations) is a suite of large-volume hydrodynamical cosmological simulations of the Virgo Consortium. These simulations are performed with the \textsc{Swift} code \citep{Schaller.etal.2024} starting from redshift $z=31$ and run to $z=0$. FLAMINGO uses some of the subgrid physics developed for the OWLS simulations \citep{Schaye.etal.2010} and used by the EAGLE, such as 
star formation, stellar mass loss, stellar feedback, supermassive black holes, and AGN feedback, but with new cooling tables \citep{Ploeckinger.Schaye.2020} and various other improvements. We refer the reader to the introductory paper \citep{Schaye.etal.2023} for more details on the FLAMINGO simulations. The gas fraction in low-redshift galaxy clusters and the galaxy stellar mass function at $z=0$ are used for calibration purposes. The FLAMINGO simulations use the Chabrier initial mass function \citep{Chabrier.etal.2003}. These calibrations are performed using machine learning \citep{Kugel.etal.2023} in contrast to traditional trial and error methods.

The fiducial FLAMINGO simulations comprise four boxes: three having 1 cGpc side length and one 2.8 cGpc. They are named L1\_m8, L1\_m9, L1\_m10, and L2p8\_m9 according to the box size (respectively 1, 1, 1, and 2.8 cGpc) and gas particle mass (respectively $1.34 \times 10^{8}~\Msun$, $1.07 \times 10^{9}~\Msun$, $8.56 \times 10^{9}~\Msun$, and $1.07 \times 10^{9}~\Msun$). L1\_m8 and L2p8\_m9 are the flagship simulations of the FLAMINGO project. Apart from the fiducial simulations, it also includes twelve variations for galaxy formation prescriptions in a 1 cGpc box: eight for different calibration data and four for different cosmologies. The simulations are stored in 79 snapshots between the redshifts $z=15$ and $z=0$. FLAMINGO uses \textsc{VELOCIraptor} \citep{Elahi.etal.2019} to implement the \textsc{FoF} algorithm on all particles (except neutrinos) in the configuration space for halo identification. Later, subhalos are identified again using the \textsc{FoF} algorithm on all the particles excluding neutrinos in the 6D phase space. The halos/subhalos are further processed using the Spherical Overdensity and Aperture Processor (SOAP, a tool developed for FLAMINGO) to compute various properties in 3D or projected apertures.

Table~\ref{tab:sim_param_tab} lists the cosmological parameters for fiducial FLAMINGO simulations and our selection cuts imposed on the sample galaxies. Note that we have imposed a higher stellar mass cut in FLAMINGO than for EAGLE and IllustrisTNG in order to consider only those galaxies which have at least ten stellar particles (down to which the observed stellar mass function is reproduced in \cite{Schaye.etal.2023}). However, this choice of the stellar mass cut does not affect our results because most to the SMGs are massive objects (see Section~\ref{sec:introduction} and Appendix~\ref{appendix:res_mass_effect}).

\section{Modeling SMG flux density}
\label{sec:smg_modeling}
Accurate modeling of the submillimeter (submm) flux densities of simulated galaxies requires full 3D dust RT calculations on-the-fly while performing simulations.  However, performing these calculations during simulations is computationally very expensive. 
Thus, post-processing the output is a more preferred and widely accepted approach, particularly for simulations without a cold ISM, such as the ones studied here. RT calculations are relevant when we have a good resolution of the ISM. Cosmological simulations, specifically with large box sizes, lack a resolved ISM. 
In such cases, thanks to the minimal dependence of (sub)mm fluxes on the ISM geometry, semi-empirical and parametric relations based on the galaxy properties are well suited \citep[see e.g.,][for some examples]{Shimizu.etal.2012,Dave.etal.2010,Hayward.etal.2021}. 
%

The simulations we use in this study are fully hydrodynamical and do not have sufficiently resolved ISM. Thus, we opt to employ computationally less expensive parametric relations. The parametric relations reported in \citet{Hayward.etal.2013} have been revised by \citet{Lovell.etal.2021} and \citet{Cochrane.etal.2023}. The general parametric form of the submm flux density at 850 $\mum$ is given by the following equation,
\begin{equation}
    S_{850}~[\mathrm{mJy}] = {a} \left(\frac{SFR}{\mathrm{100~\MsunYr}}\right)^b \left(\frac{M_{\mathrm{dust}}}{\mathrm{10^{8}~\Msun}}\right)^c,
    \label{eqn:s850_relation}
\end{equation}
where SFR is the instantaneous star formation rate, and $M_{\rm dust}$ is the dust mass. The indices $a$, $b$, and $c$ are constants coming from fitting to the results of radiative transfer calculations. In Table~\ref{tab:s850_parameters}, we list the values of these constants derived from RT calculations on different simulations. For the submm flux density calculation, the instantaneous SFR is readily available for hydrodynamical simulations and the dust mass can be obtained from the metal mass assuming a constant dust-to-metal ratio.
%
\begin{table*}
    \centering
    \caption{The list of fitting parameters derived from radiative transfer calculations on different simulations for the S$_{850}$ flux density estimation using equation~(\ref{eqn:s850_relation}). \cite{Hayward.etal.2013} did not report uncertainties in individual parameter. Instead, they found a 0.13 dex scatter between the flux densities obtained from their relation and their radiative transfer calculations.}
    \begin{tabular}{|llllll|}
    \hline
    $a$ & $b$ & $c$ & Simulations/data used & References & Name \\
    \hline
    0.81 & 0.43 & 0.54 & Idealized merging and isolated galaxies & \citet{Hayward.etal.2013} & H13 \\
    0.58$\pm$0.002 & 0.51$\pm$0.002 & 0.49$\pm$0.003 & SIMBA cosmological simulations & \citet{Lovell.etal.2021} & L21 \\
    0.55$\pm$0.04 & 0.50$\pm$0.09 & 0.51$\pm$0.06 & FIRE-2 zoom-in simulations & \citet{Cochrane.etal.2023} & C23 \\
    \hline
    \end{tabular}
    \label{tab:s850_parameters}
\end{table*}

\subsection{(Sub)mm population synthesis}
\label{sec:submm_pop_synth}
In this section, we describe the implementation of parametric models for (sub)mm population synthesis. For this, we first compute the SFR and $M_{\rm dust}$ properties of each subhalo (i.e. galaxy). 
For the SFR calculation, a spherical aperture of 30 physical kpc centered on the subhalo is used, and the instantaneous SFRs of each star-forming gas cell within the aperture are summed. For dust mass we assume a fixed dust-to-metal (DTM) mass ratio of 0.4 following observational estimates form \cite{Dwek.1998} and \cite{James.etal.2002}. Further, we consider only cold star-forming gas cells for the metal mass calculation because thermal sputtering and collisions in the hot gas cells destroy dust grains \citep{Draine.Salpeter.1979, McKinnon.etal.2016, Popping.etal.2017}. By virtue of the star formation criterion, all the gas cells that are forming stars are cold and dense for the metal mass calculation. The sum of the metal mass in all the cold star-forming gas cells returns the total metal mass of the galaxy, which is converted into a dust mass using DTM $=0.4$. The exact DTM ratio does not affect the measured flux densities significantly. Using DTM $=0.5$ would increase the flux densities by only $11-13\%$ for all the models listed in Table~\ref{tab:s850_parameters}.

\subsection{Choice of the parametric model}
\label{sec:choice_of_model}
To choose the parametric model for (sub)mm galaxies population synthesis for further study, we test the predictions of the models listed in Table~\ref{tab:s850_parameters} for the EAGLE cosmological simulations, and for observational data in the following subsections.

\subsubsection{Testing parametric models on the EAGLE simulation}
\label{sec:model_test_for_eagle}
First, we use the parametric relations listed in Table~\ref{tab:s850_parameters} on the EAGLE galaxies, where RT calculations are already available in the SCUBA 850 $\mum$ band for comparison. \cite{Camps.etal.2018} estimated the photometric fluxes for the EAGLE galaxies using the dust radiative transfer code \textsc{skirt} \citep{Camps.Baes.2015}.
We download the EAGLE data from their database using the SQL query. We compile information on the star-forming gas mass, instantaneous star formation rate, star-forming gas metallicity, and SCUBA 850 flux density for all the snapshots available for the RefL0100N1504 simulation for galaxies with stellar mass greater than $\rm 10^{8.5}~\Msun$.

\begin{figure}
    \centering
    \includegraphics[width=\columnwidth]{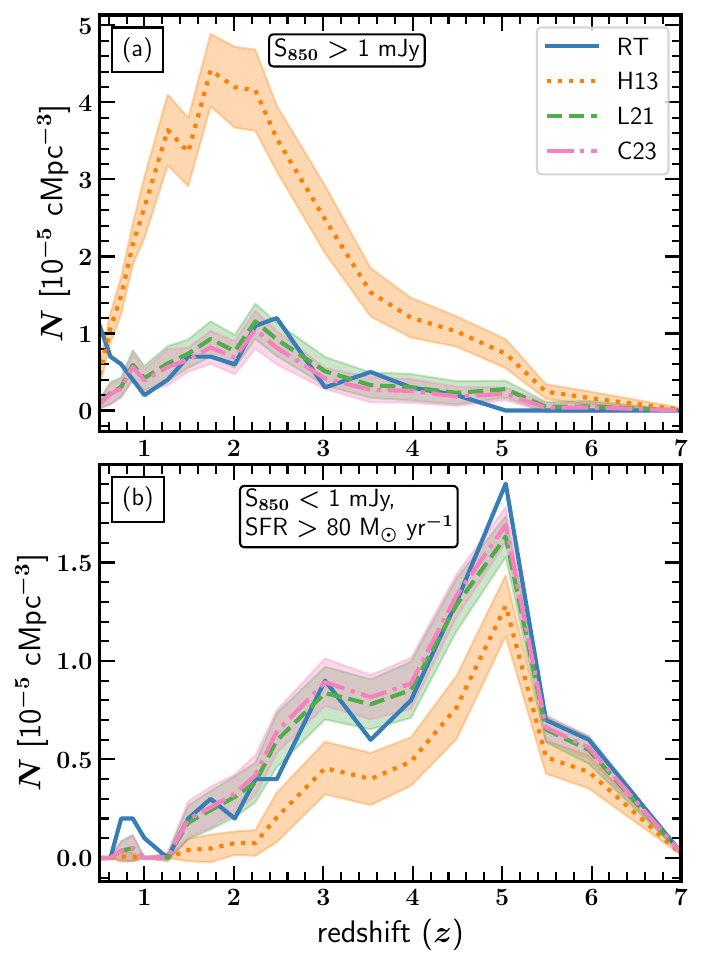}
    \caption{Comparison of the S$_{850}$ flux densities of the EAGLE galaxies calculated using radiative transfer or the parametric relation given in equation~(\ref{eqn:s850_relation}) with the parameter values listed in Table~\ref{tab:s850_parameters}. Here, solid blue color shows radiative transfer (RT) results, dotted orange shows the results for the \citet{Hayward.etal.2013} relation (H13), dashed green the \citet{Lovell.etal.2021} relation (L21), and dash-dotted pink the \citet{Cochrane.etal.2023} relation (C23). The shaded regions represent 1$\sigma$ uncertainties. Panels (a) and (b) show the evolution of the comoving number density for, respectively submm-bright galaxies (S$_{850} > 1$ mJy), and submm-faint star-forming galaxies (S$_{850} < 1$ mJy, SFR $> 80~\Msun/yr$). 
    The L21 and C23 relations exhibit very similar results and are consistent with the RT calculations. However, the H13 relation over-estimates the submm flux densities compared to RT.}
    \label{fig:RT_Hay13_Lov21_comparison}
\end{figure}

Similar to the RT calculations for EAGLE \cite[see][]{Camps.etal.2018}, we assume a fixed dust-to-metal ratio of 0.3 for this test and measure the S$_{850}$ flux density for three sets of parameters listed in Table~\ref{tab:s850_parameters} and compare their results in Fig.~\ref{fig:RT_Hay13_Lov21_comparison}. To incorporate the effect of uncertainties in the parameters of the fitted parametric relations, we apply a Gaussian scatter of 0.13 dex on the modeled submm flux densities. 
We generate 100 realizations of flux densities by randomly applying this Gaussian scatter. In Fig.~\ref{fig:RT_Hay13_Lov21_comparison}, the orange color shows the \citet{Hayward.etal.2013} relation (H13), green shows the \citet{Lovell.etal.2021} relation (L21), and pink shows the \citet{Cochrane.etal.2023} relation (C23). The RT calculations from the EAGLE data are shown in blue. The shaded regions around the measurements from three parametric relations represent 1$\sigma$ uncertainties around the mean of 100 realizations of the flux densities.

In Fig.~\ref{fig:RT_Hay13_Lov21_comparison}(a), we compare the number density of submm-bright galaxies with S$_{850}>$ 1 mJy as a function of redshift. We find that the results from L21 and C23 are comparable to each other and also consistent with the RT calculations. On the other hand, the distribution from the H13 relation overpredicts the number of submm-bright galaxies at all the redshifts reaching about 5 times the number density from the radiative transfer calculations at $z \approx 2$. Similarly, in Fig.~\ref{fig:RT_Hay13_Lov21_comparison}(b), we compare the number density of galaxies with S$_{850}<$ 1 mJy, but forming stars at a rate greater than 80 $\MsunYr$. Again, we find good agreement between the results of using the L21 and C23 models as well as their consistency with RT calculations. In contrast, the H13 relation under-predicts submm-faint star-forming galaxies.

A comparison of the parameter values in Table~\ref{tab:s850_parameters} shows that the main difference between the models is the normalization factor. This impacts directly on the number of galaxies exceeding the flux density of S$_{850} >$~1~mJy as seen in Fig.~\ref{fig:RT_Hay13_Lov21_comparison}(a). As expected, this also influences the number of submm-faint star-forming galaxies as shown Fig.~\ref{fig:RT_Hay13_Lov21_comparison}(b). 

Our tests and findings with the EAGLE simulation show that the S$_{850}$ predictions using the L21 and C23 relations are remarkably closer to the RT calculation compared to the H13 relation. \citet{Lovell.etal.2021} obtain their parametric relation from the SIMBA cosmological simulations and \citet{Cochrane.etal.2023} from the FIRE-2 zoom-in cosmological simulations, while \citet{Hayward.etal.2013} use idealized simulations of isolated and merging galaxies. 
Each of these studies employs different cosmological simulations, varying in spatial and mass resolution as well as subgrid physics.
In addition, these studies employ different RT codes in their analyses, which could also influence their findings. The H13 used \textsc{sunrise} \citep{Jonsson.etal.2010}, L21 used \textsc{powderday} \citep{Narayanan.etal.2021}, C23 used \textsc{skirt} \citep{Camps.Baes.2015}, and the EAGLE database provides radiative transfer calculations using \textsc{skirt}.

\subsubsection{Testing parametric models on observational data}
\label{sec:model_test_for_observation}
After testing the parametric relations listed in Table~\ref{tab:s850_parameters} using the EAGLE cosmological simulations, we evaluate their predictions for real observational data. For this purpose, we 
searched for the archival data of (sub)mm galaxies where measured star formation rates and dust masses are available. We use data from
\cite{da_Cunha.etal.2015}, \cite{Dudzeviciute.etal.2020}, and \cite{Hyun.etal.2023}. The complete sample comprises 892 submm galaxies: 99 from ALMA at $870~\mum$ in the Extended Chandra Deep Field South \citep[ECDF-S:][]{da_Cunha.etal.2015}, 707 from ALMA at $870~\mum$ in UKIRT Infrared Deep
Sky Survey (UKIDSS) Ultra Deep Survey \citep[UKIDSS-UDS:][]{Dudzeviciute.etal.2020}, and 86 from JCMT at $850~\mum$ in the JWST Time-Domain Field (JWST-TDF) near the North Ecliptic Pole (NEP) \citep{Hyun.etal.2023}. All these data use the MAGPHYS galaxy SED-fitting code \citep{da_Cunha.etal.2008, Battisti.etal.2019} to estimate star formation rates and dust masses. We compiled observational flux densities, star formation rates, and dust masses from above mentioned datasets for our analysis. The flux densities at $870~\mum$ are scaled to $850~\mum$ by multiplying by 1.09 assuming the modified black body relation following equation~(\ref{eqn:flux_conversion}) in Section~\ref{sec:source_number_counts}.

\begin{figure}
    \centering
    \includegraphics[width=\columnwidth]{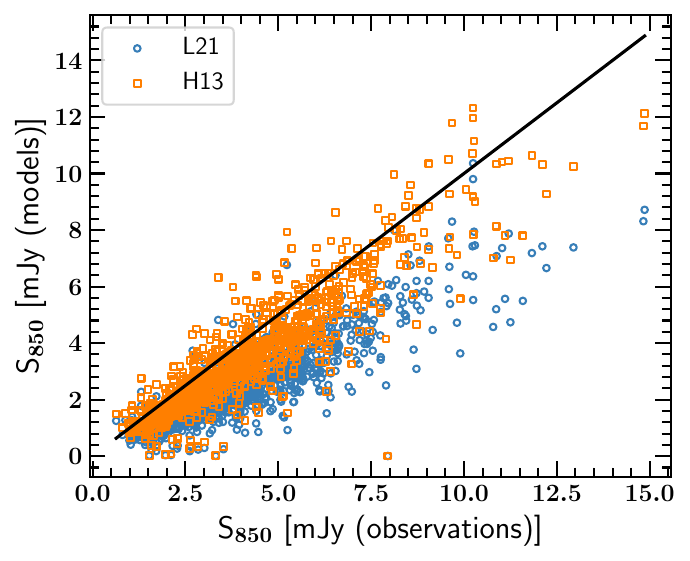}
    \caption{Comparison of observed flux densities with flux densities predicted using the parametric relations listed in Table~\ref{tab:s850_parameters}, and observed star formation rates and dust masses. The circles in blue and boxes in orange represent, respectively, the L21 and H13 relations. The black solid line is line of equality for visual guidance. It is clear that H13 better reproduces observed fluxes than L21 (and C23). See Section~\ref{sec:model_test_for_observation} for more details.}
    \label{fig:models_vs_observations}
\end{figure}

In Fig.~\ref{fig:models_vs_observations}, we compare $850~\mum$ flux densities obtained using the parametric relations listed in Table~\ref{tab:s850_parameters} with the observed flux densities at $850~\mum$. Since the L21 and C23 relations give very similar results (as seen for the EAGLE simulation in Section~\ref{sec:model_test_for_eagle}), we show only the comparisons of H13 and L21 for clarity. Here, blue circles and orange boxes show L21 and H13 predictions, respectively. Observed flux densities are shown on the x-axis and modeled flux densities are shown on the y-axis. The black solid line represents one-to-one relation for visual guidance. It is clear that the H13 model better represents the observations than the L21 model for the whole flux density range. The L21 model under-predict the flux densities for brighter galaxies. This is the first analysis in which we have tested and compared the predictability of these models on observed fluxes.

To conclude our test on observational data, we find that the H13 relation better reproduces observed flux densities for given observed star formation rates and dust masses. However, L21 under-predicts for brighter galaxies. 
Our analysis shows that the L21 and C23 models predict similar flux densities, but lower than the H13 model. The L21 model works better for the EAGLE simulation, whereas H13 works better for observations. Hence, for the sake of completeness, we show our results using the L21 and H13 relations unless explicitly stated otherwise.

\section{Results}
\label{sec:results}
In this section, we discuss our results from the TNG100, TNG300, and FLAMINGO simulations. We particularly focus on the new large-volume FLAMINGO simulations. For clarity, in the main text, we show results for the fiducial FLAMINGO simulation L1\_m8, and refer to it as FLAMINGO unless explicitly stated otherwise. The FLAMINGO simulations with different box sizes and resolutions, different galaxy formation prescriptions, and cosmologies are discussed briefly in the appendices (see Appendices~\ref{appendix:gal_form_pres_effect} and \ref{appendix:gal_form_cosmo_effect}). In this section, we present the redshift distributions, source number counts, contribution to star formation rate density, star formation rates, and flux density functions. We also compare our findings with the available observational data. 

\subsection{Redshift distribution}
\label{sec:redshift_distribution}

\begin{figure}
    \centering
    \includegraphics[width=\columnwidth]{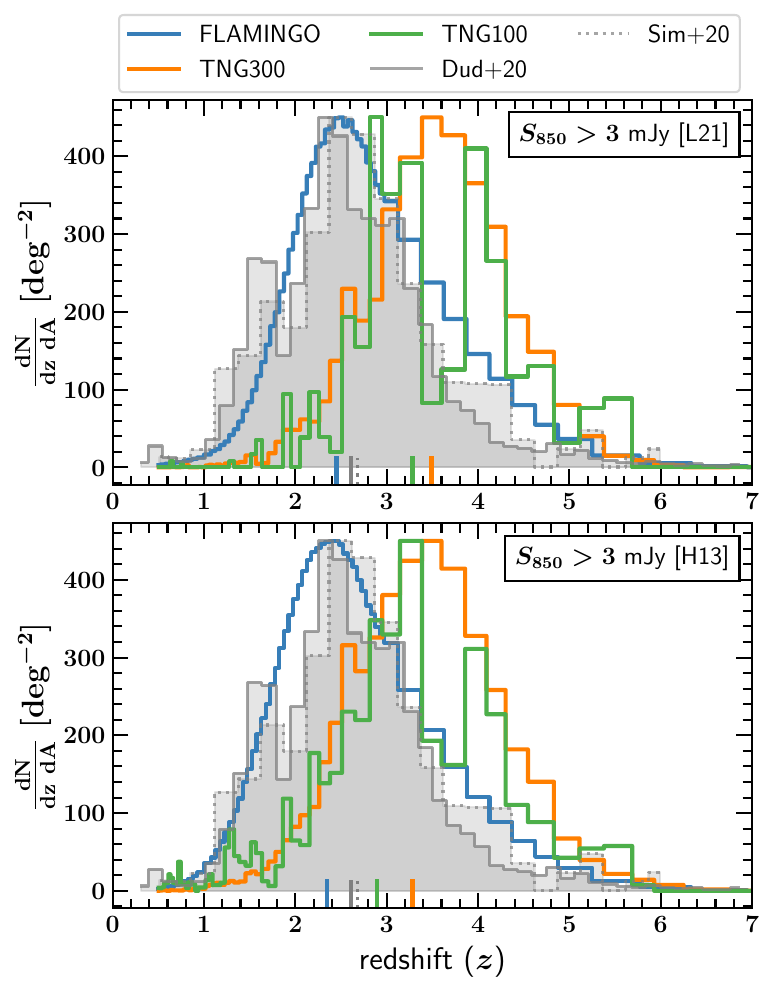}
    \caption{Redshift distribution of submm-bright galaxies with S$_{850}>3$~mJy in the FLAMINGO (blue), TNG300 (orange), TNG100 (green) simulations. gray color histograms represent recent observational estimates from \cite{Dudzeviciute.etal.2020} [Dud+20], and \cite{Simpson.etal.2020} [Sim+20]. All the distributions have been re-scaled to match the height of the \cite{Dudzeviciute.etal.2020} distribution. The small bars at the bottom of each panel indicate the medians of the respective distributions. Top panel shows predictions assuming the L21 relation, while the bottom panel is for the H13 relation. Quantitatively, FLAMINGO better reproduces the observed distribution for the H13 relation (see Section~\ref{sec:redshift_distribution} for more details).}
    \label{fig:redshift_dist_all_sim}
\end{figure}

The redshift distribution of submm galaxies provides relevant information on their assembly and peak activity. 
In this subsection, we focus on the shape of our predicted distributions. The submm source number counts are discussed in the next subsection. In Fig.~\ref{fig:redshift_dist_all_sim}, we present the redshift distributions of submm-bright galaxies (S$_{850} > 3$ mJy) for the FLAMINGO, TNG300, and TNG100 simulations in, respectively, blue, orange, and green colors. The top panel shows predictions using the L21 relation, and the bottom panel displays predictions assuming the H13 relation. The gray colors show the recent observational redshift distribution from \cite{Dudzeviciute.etal.2020} and \cite{Simpson.etal.2020}. Note that we have normalized all the distributions by the projected sky area and the redshift bin-sizes. At the bottom of each panel, we indicate the median values of each distribution with their respective colors using small vertical bars.

For comparison, we have rescaled all the distributions to match the height of observed distribution reported in \cite{Dudzeviciute.etal.2020}. This rescaling of the FLAMINGO, TNG300, TNG100 distributions, respectively, requires factors 3.7, 17.8 and 25.5 for the L21 relation, and 0.8, 6.8 and 8.7 for the H13 relation. The \cite{Simpson.etal.2020} distribution requires a scaling factor 4.7 to match the height of \cite{Dudzeviciute.etal.2020}. We remind the reader that \cite{Dudzeviciute.etal.2020} carried out ALMA follow-up of 716 submm galaxies with S$_{850} > 3.6$ mJy in a 0.96 deg$^{2}$ sky area. Their final sample comprises 707 submm galaxies with S$_{870} > 0.6$ mJy. On the other hand, \cite{Simpson.etal.2020} carried out ALMA follow-up of 183 submm galaxies with S$_{850} > 6.2$ mJy in 1.6 deg$^{2}$ area. Their final sample comprises 182 submm galaxies with S$_{870} > 0.7$ mJy. The higher flux density selection threshold employed by \cite{Simpson.etal.2020} compared to \cite{Dudzeviciute.etal.2020} requires scaling their distribution by a factor of 4.7 to align it with the latter.

For the L21 relation, TNG100 and TNG300 show very similar distributions with a small irregularity in TNG100 at $z \approx 3.5$ due to its relatively small box size for statistical sampling of bright SMGs. As compared to the observed distributions, both TNG100 and TNG300 peak at higher redshifts. The median redshifts of TNG100 and TNG300 are $z_{med} = 3.28 \pm 0.051$ and $3.49 \pm 0.001$, respectively. On the other hand, the FLAMINGO simulation shows remarkable similarity with the shape of observed distributions having median redshift $z_{med}=2.45\pm0.001$. The median redshift for \cite{Dudzeviciute.etal.2020} is $z_{med}=2.61\pm0.08$ and the median redshift for \cite{Simpson.etal.2020} is $z_{med}=2.68\pm0.06$. In addition, the low- and high-redshift tails of the FLAMINGO simulation nicely trace the observed pattern.  The small difference seen between the median redshifts of FLAMINGO and the observational data may be due to the effect of sample selection. There are evidences that the redshift distributions for galaxies selected at longer wavelengths and/or brighter flux density tend to peak at higher redshift \citep[e.g.,][]{Pope.etal.2005, Brisbin.etal.2017, Lagos.etal.2020, Reuter.etal.2020}. We will discuss this further later in this section.

The TNG100 and TNG300 redshift distributions are very similar if instead of the L21 we use the H13 relation as shown in the bottom panel of Fig.~\ref{fig:redshift_dist_all_sim}, again with a small irregularity at $z \approx 3.5$ in TNG100. However, compared with the predictions using the L21 relation, the predictions using H13 relation move slightly towards lower redshifts. The median redshifts for TNG100 and TNG300 drift to $z_{med}=2.90 \pm 0.048$ and $3.28 \pm 0.112$ respectively. Still, the overall distributions of TNG100 and TNG300 are far from the observed distributions. On the other hand, the median redshift for FLAMINGO changes to $2.35 \pm 0.001$. Though FLAMINGO shows a small shift towards low-redshift, its predictions are matching the observed distributions quite well. 

\begin{figure}
    \centering
    \includegraphics[width=\columnwidth]{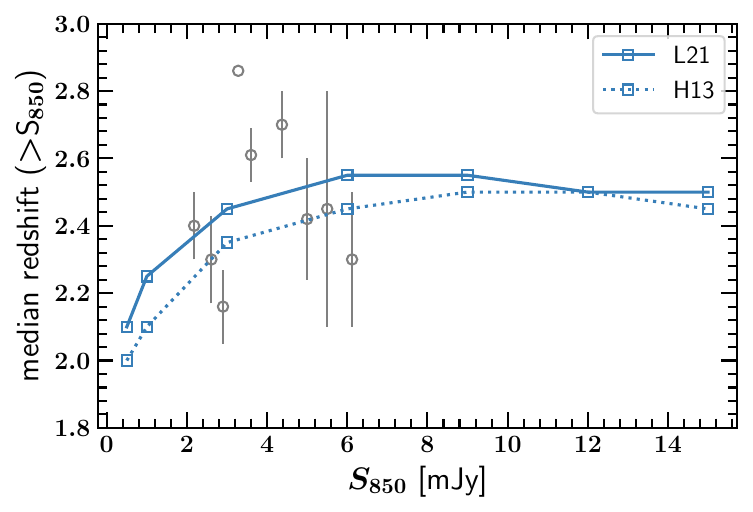}
    \caption{The median of the redshift distributions of submm galaxies as a function of the flux density cut in FLAMINGO simulation. The solid blue curve represents the prediction using the L21 relation, and the dotted blue curve shows the prediction assuming the H13 relation. Gray color points represent observational estimates obtained from \cite{Chapman.etal.2005, Pope.etal.2005, Wardlow.etal.2011, Casey.etal.2013, da_Cunha.etal.2015, Danielson.etal.2017, Miettinen.etal.2017, Cowie.etal.2018, Dudzeviciute.etal.2020}. Both the models show consistency with observations, and predict negligible evolution in the median redshift for S$_{850} > 8$ mJy.}
    \label{fig:median_redshift}
\end{figure}

As pointed out earlier, the median of the redshift distribution moves towards higher redshifts when we select brighter galaxies \citep{Pope.etal.2005, Brisbin.etal.2017} or observe at longer wavelengths \citep{Lagos.etal.2020, Reuter.etal.2020}. The statistically representative volume of FLAMINGO (in particular) provides us with the opportunity to compare these observed trends with predicted distributions from parametric relations. To investigate the effect of the flux density cuts on the median values of the redshift distributions, we plot the median of redshifts as a function of the S$_{850}$ flux density cut in Fig.~\ref{fig:median_redshift} for FLAMINGO. The solid and dotted blue curves represent, respectively, predictions using the L21 and H13 relations. For comparison, we over-plot the observed data points in gray color compiled from various literature \citep{Chapman.etal.2005, Pope.etal.2005, Wardlow.etal.2011, Casey.etal.2013, da_Cunha.etal.2015, Danielson.etal.2017, Miettinen.etal.2017, Cowie.etal.2018, Dudzeviciute.etal.2020}. For both the relations, there is a clear trend of increasing median redshift as we choose a larger flux density cut for selecting brighter galaxies. However, both trends flatten towards higher flux density cuts. Due to the limited observational data, we cannot verify this flattening. Future surveys with large sky coverage will allow us to test these predictions of the FLAMINGO simulation. Additionally, as seen in Fig.~\ref{fig:redshift_dist_all_sim}, the H13 relation predicts a lower/equal median redshift than the L21 relation at any given flux density cut. Given the range of uncertainties in the observations, both the L21 and H13 relations predict observed median redshift trend quite well.

\subsection{Source number counts}
\label{sec:source_number_counts}
\begin{figure}
    \centering
    \includegraphics[width=\columnwidth]{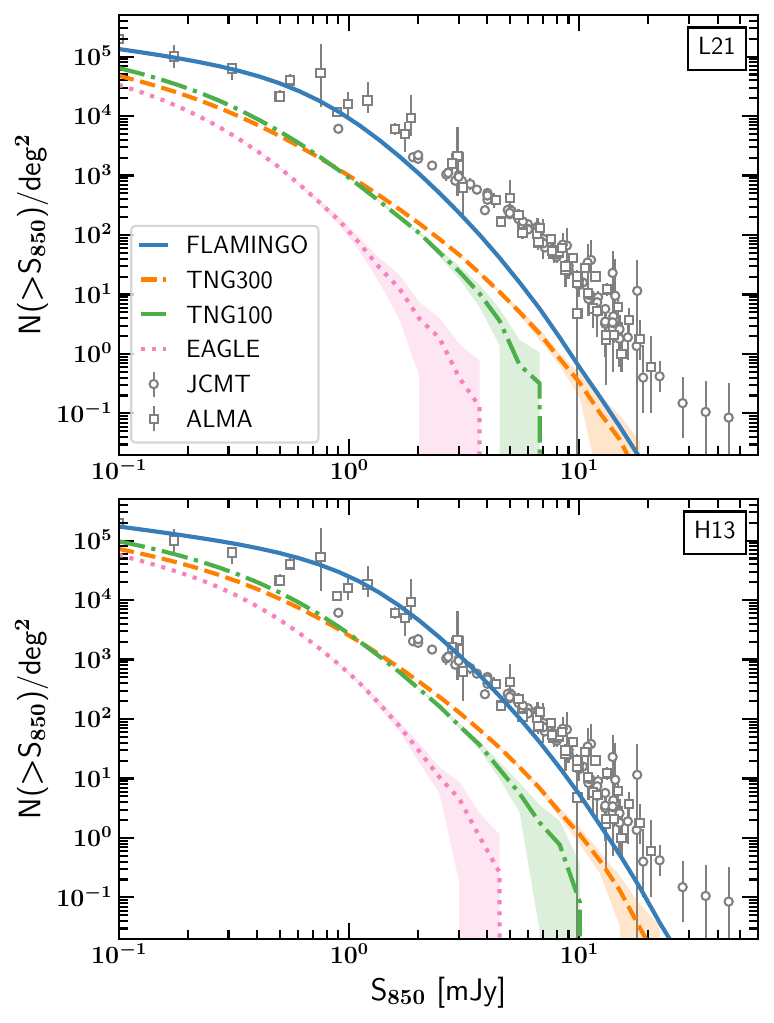}
    \caption{Source number counts of submm galaxies in the FLAMINGO (solid blue), TNG300 (dashed orange), TNG100 (dash-dotted green), and EAGLE (dotted pink) simulations. Shaded regions show 1$\sigma$ errors for 100 realizations of the modeled submm galaxy population, as discussed in Section~\ref{sec:model_test_for_eagle}. The top panel shows predictions using the L21 relation while the bottom panel uses the H13 relation. For comparison, we have overlaid observed source number counts in gray color, which are grouped as JCMT \citep{Casey.etal.2013, Geach.etal.2017, Zavala.etal.2017, Simpson.etal.2019, Shim.etal.2020, Garratt.etal.2023, Hyun.etal.2023, Zeng.etal.2024} and ALMA \citep{Hatsukade.etal.2013, Karim.etal.2013, Simpson.etal.2015, Fujimoto.etal.2016, Hatsukade.etal.2018, Stach.etal.2018, Simpson.etal.2020} according to the observing facility used. We have scaled all the flux densities of ALMA bands to the S$_{850}$ band assuming a modified black body. The FLAMINGO predicts the observed number counts quite well for the H13 relation. See Section~\ref{sec:source_number_counts} for more detail.}
    \label{fig:number_counts_all_sim}
\end{figure}

Number counts represent the cumulative number count of (sub)mm sources per unit of projected area as a function of flux density.
To obtain the source number counts for simulations, we use the method described in \citet{Hayward.etal.2013}. The cumulative source number counts per unit deg$^{2}$ at any given flux density $S_\lambda$ is given by the following expression:

\begin{equation}
    \frac{N (>S_\lambda)}{\rm [deg^{2}]} = \left(\frac{\pi}{180}\right)^{2} \int \frac{dN(>S_\lambda)}{dV}\frac{dV}{d\Omega dz} dz,
    \label{eqn:number_counts}
\end{equation}
where the comoving volume element per unit solid angle ($d\Omega$) per unit redshift interval ($dz$) is given as
\begin{equation}
    \frac{dV}{d\Omega dz} = \frac{c}{H(z)} D_{C}^{2}(z),
\end{equation}
where the comoving distance to redshift $z$ is
\begin{equation}
    D_{C}(z) = c \int \frac{dz}{H(z)},
    \label{eqn:angular_distance}
\end{equation}
and the redshift-dependent Hubble parameter in a flat universe is
\begin{equation}
    H(z) = H_{0} \sqrt{\Omega_{\mathrm{M}}(1+z)^{3}+\Omega_{\Lambda}}.
\end{equation}
In equation~(\ref{eqn:number_counts}), $dN(>S_\lambda)/dV$ for any simulation snapshot is computed by diving the total number of sources brighter than $S_\lambda$ by the comoving volume of the simulation box. Then we perform a linear interpolation of $dN(>S_\lambda)/dV$ between the snapshots when evaluating the integral in equation~(\ref{eqn:number_counts}).

In Fig~\ref{fig:number_counts_all_sim}, we show predicted source number counts of submm galaxies in the FLAMINGO (solid blue), TNG300 (dashed orange), TNG100 (dash-dotted green), and EAGLE (dotted pink) simulations. The shaded regions represent the 1$\sigma$ scatter around the mean for 100 realizations as discussed for EAGLE in Section~\ref{sec:model_test_for_eagle}. The top panel shows the source number counts obtained using the L21 relation, and the bottom panel shows the same for the H13 relation. The gray color open circles and open boxes represent observed source number counts compiled from the literature. For clarity, we have grouped the observed data as JCMT \citep{Casey.etal.2013, Geach.etal.2017, Zavala.etal.2017, Simpson.etal.2019, Shim.etal.2020, Garratt.etal.2023, Hyun.etal.2023, Zeng.etal.2024} and ALMA \citep{Hatsukade.etal.2013, Karim.etal.2013, Simpson.etal.2015, Fujimoto.etal.2016, Hatsukade.etal.2018, Stach.etal.2018, Simpson.etal.2020} depending on the observing facility used for these observation. Note that the observed flux density in any (sub)mm band $S_{X}$ (or $S_{\lambda}$) is scaled to match the S$_{850}$ band assuming the following relation used in various publications \citep[e.g.,][]{Dunne.etal.2001, Simpson.etal.2019, Zeng.etal.2024}.

\begin{equation}
    \frac{S_{850}}{S_{X}} = \left(\frac{\nu_{850}}{\nu_{X}}\right)^{\beta} \frac{B(\nu_{850}, T)}{B(\nu_{X}, T)} \approx \left(\frac{\nu_{850}}{\nu_{X}}\right)^{\beta + 2} = \left(\frac{\lambda_{X}~[\mum]}{850}\right)^{\beta + 2}.
    \label{eqn:flux_conversion}
\end{equation}
This simple flux density conversion relation is the result of (sub)mm wavelengths being in the Rayleigh-Jeans region of
the Planck function ($B(\nu, T)$). Assuming $\beta=1.8$ \citep{planck_collaboration.2011}, we calculate conversion factors 5.03 for $1.3~\mm$, 3.71 for $1.2~\mm$, 2.66 for $1.1~\mm$, and 1.09 for 870~$\mum$. 

None of the simulations reproduce the observed source number counts when using the L21 relation (see the top panel of Fig.~\ref{fig:number_counts_all_sim}). 
At 1 mJy, observational number counts are a factor of $\sim$100 higher than the EAGLE predictions. Although TNG100 and TNG300 show some improvement as compared to EAGLE, they are far from the observations. Toward the fainter end, the small drop in TNG300 compared to TNG100 is the result of the lower mass resolution of TNG300. There is an excess at bright end in TNG300 as compared to TNG100, showing the importance of a large box size to statistically represent the bright end of number counts. At 1 mJy, the observations are about an order of magnitude higher than the TNG predictions. In contrast, FLAMINGO reproduces the observed number counts for S$_{850} < 2$ mJy relatively well. 

All the source number count distributions shift upward and rightward when using the H13 relation (see the bottom panel of Fig.~\ref{fig:number_counts_all_sim}). Although EAGLE, TNG100, and TNG300 show an increase in the source number counts, they remain very discrepant with the observations. On the other hand, FLAMINGO reproduces the observed number counts quite well when using H13 relation. The small excess of observed submm-bright sources relative to FLAMINGO is possibly the effect of multiple source blending as reported in resolved interferometric observations \citep[see,][]{Chen.atal.2013, Hodge.etal.2013, Karim.etal.2013}. 

Overall, the FLAMINGO simulation simultaneously reproduces the redshift distribution and source number counts of SMGs when using H13 relation. Also, the H13 relation performs better than L21 when applied to observations (see Fig.~\ref{fig:models_vs_observations}). Hence, it remains valuable for statistical comparison of the observed submm universe with the simulated universe in FLAMINGO when modeled with the H13 relation.


\subsection{Contribution to the cosmic SFRD}
\label{sec:sfrd_contribution}
\begin{figure*}
    \centering
    \includegraphics[width=\textwidth]{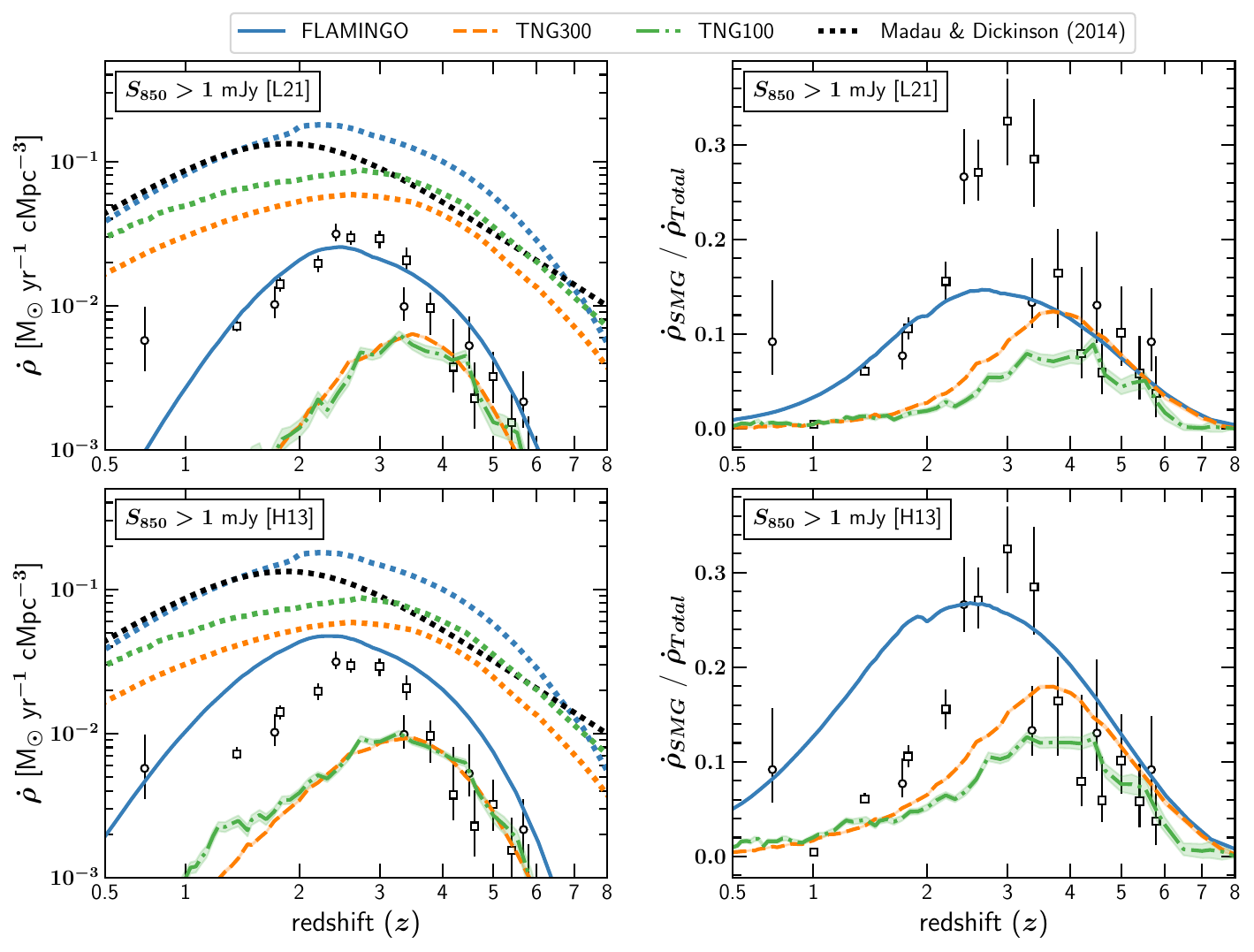}
    \caption{\textit{Left column}: Contribution of submm galaxies with S$_{850}>1$ mJy to the cosmic star formation rate density in the FLAMINGO (solid blue) TNG300 (dashed orange), and TNG100 (dash-dotted green) simulations. Dotted curves represent the total SFRD in corresponding colors. The dotted black color show the \cite{Madau.Dickinson.2014} fit to observations. The top panel shows predictions using the L21 relation, while the bottom panel using the H13 relation. The shaded regions around SFRD of submm galaxies are 1$\sigma$ uncertainty for 100 realizations. For comparison, we show submillimeter observations from \cite{Swinbank.etal.2014} and \cite{Dudzeviciute.etal.2020} using, respectively, open circles and boxes. \textit{Right column}: The ratio of submm to total SFRD correspond to the left column. We use \cite{Madau.Dickinson.2014} cosmic SFRD for estimating observational ratio. At peak activity, submm contribution in the FLAMINGO is about 27\% for the H13 Model.}
    \label{fig:sfr_dens_all_sim}
\end{figure*}

(Sub)mm galaxies being among the most star-forming objects in the universe, we would like to know their contribution to the total cosmic star formation rate density (SFRD). To investigate this in the FLAMINGO, TNG300, and TNG100 simulations, we plot their submm SFRDs for submm-bright galaxies with S$_{850}>1$ mJy in the left column panels of Fig.~\ref{fig:sfr_dens_all_sim}, using solid blue, dashed orange, and dash-dotted green curves, respectively. Their total SFRDs are shown with dotted curves in respective colors. For comparison, we have also over-plotted the \cite{Madau.Dickinson.2014} fit to observations with dotted black color. The shaded regions around the SFRD of submm galaxies represent 1$\sigma$ scatter for 100 realization as discussed in Section~\ref{sec:model_test_for_eagle}. The top and bottom panels display the predictions for the L21 and H13 relations, respectively. For the total SFRD calculation, we have simply added the star formation rates of all the subhalos identified in a given snapshot of the respective simulation. Using the FLAMINGO simulations, we verified that our total SFRD is identical to the SFRD measured on-the-fly using all the gas particles during simulation \citep[as reported in ][]{Schaye.etal.2023}. The open circles and boxes represent observational estimates for the SFRD of submm galaxies brighter than 1 mJy from \cite{Swinbank.etal.2014} and \cite{Dudzeviciute.etal.2020}, respectively. The total SFRD and the ratio of submm to total SFRD of FLAMINGO show a small kink at $z\approx2$. This kink is only visible in the high-resolution model and absent in other FLAMINGO models (see, Fig~\ref{appendix:res_mass_effect}, \ref{appendix:gal_form_pres_effect}, ~\ref{appendix:gal_form_cosmo_effect}). Presently, we do not have an explanation for this kink \citep[see,][]{Schaye.etal.2023}.

From the left column of Fig.~\ref{fig:sfr_dens_all_sim} it is clear that FLAMINGO predicts significantly higher contribution of submm-bright galaxies to the cosmic SFRD than TNG100 and TNG300. There is also a significant difference between the peak SFRD of the submm galaxies; FLAMINGO peaks at a lower redshift ($z \approx 2.6$) than IllustrisTNG ($z \approx 3.7$). A similar shift is also evident for the total cosmic SFRD, which indeed is the reason for the shift in the redshift distribution of IllustrisTNG submm galaxies towards higher redshifts, as seen in Fig.~\ref{fig:redshift_dist_all_sim}. Since the H13 relation predicts higher submm flux densities than the L21 relation, we see a larger contribution of submm galaxies when using the H13 relation as can be perceived from the comparison of the top and bottom panels. Additionally, the observed maximum contribution of submm galaxies to the cosmic SFRD is better reproduced in FLAMINGO simulations when using the H13 relation (see the bottom-right panel in Fig.~\ref{fig:sfr_dens_all_sim}).


From the left panels of Fig.~\ref{fig:sfr_dens_all_sim}, it seems that the contributions of submm galaxies in TNG100 and TNG300 simulations are consistent with each other. However, their total cosmic SFRDs differ significantly. 
The FLAMINGO simulations were not calibrated to the cosmic SFRD or any other $z>0$ galaxy observations \citep{Schaye.etal.2023}, yet they show remarkable consistency with observation, particularly for $z<2$. On the other hand, the IllustrisTNG simulations were calibrated to the cosmic SFRD for $z<10$ \citep{Pillepich.etal.2018}, yet they disagree with the observations. For instance, TNG100 is consistent with observation for $z > 3$, but TNG300 has a lower SFRD in the whole redshift range. Therefore, to quantify the contribution of submm galaxies to the cosmic SFRD, we calculate the ratio of the submm SFRD to the total SFRD as shown in corresponding right column panels of Fig.~\ref{fig:sfr_dens_all_sim}. 
The maximum submm contribution to cosmic SFRD in the FLAMINGO simulation reaches to $\approx 15\%$ for the L21 relation and $\approx 27\%$ for the H13 relation at redshift $z=2.6$. However, the highest contribution of submm galaxies to the cosmic SFRD in TNG300 reaches only $\approx 12\%$ for the L21 relation and $\approx 18\%$ for the H13 relation at redshift $z=3.7$ which is about 2/3 of the FLAMINGO prediction. Further, the maximum submm contribution to the cosmic SFRD in FLAMINGO reasonably matches with observational estimates when using the H13 relation. The TNG100 predicts an even lower contribution than TNG300. This comparison between TNG300 and TNG100 highlights the importance of a large box size for the statistical modeling of the submm galaxy population in cosmological simulations. In Appendix~\ref{appendix:res_mass_effect}, we show the effect of box size for the FLAMINGO simulations, and confirm that a 1~Gpc box is large enough to robustly model submm galaxy population. 

\subsection{Star formation rates and stellar masses}
\label{sec:sfr_stellar_mass}

\begin{figure*}
    \centering
    \includegraphics[width=\textwidth]{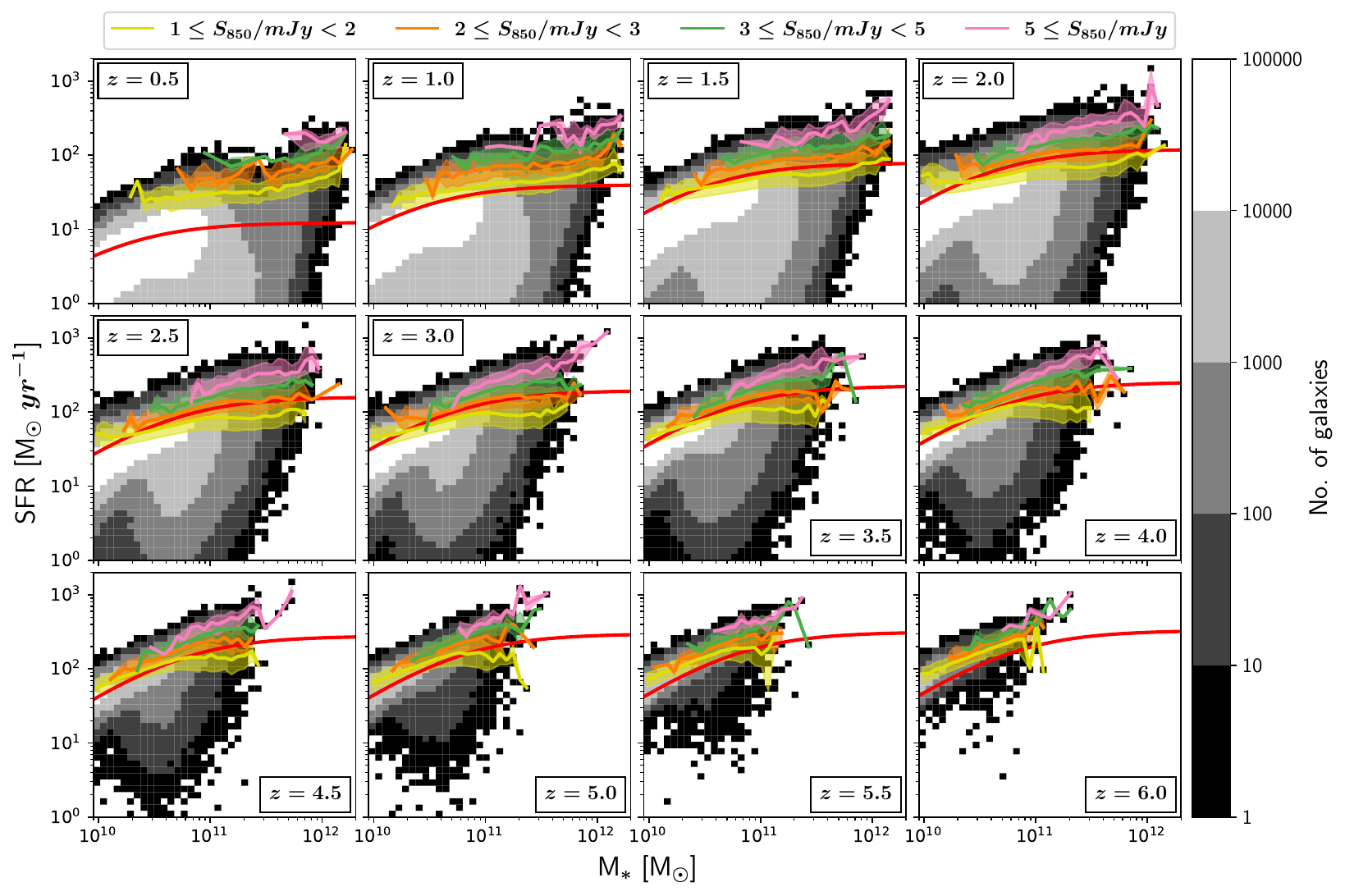}
    \caption{The evolution of the median star formation rate $-$ stellar mass relation in the FLAMINGO simulation from $z=6$ to $z=0.5$ for SMGs grouped into four flux density ranges shown with yellow ($\rm 1 \le S_{850}/mJy < 2$), orange ($\rm 2 \le S_{850}/mJy < 3$), green ($\rm 3 \le S_{850}/mJy < 5$), and pink ($\rm S_{850} \ge 5~mJy$) color curves. Their respective shaded regions represent 16th to 84th percentiles. The background gray color maps show the distribution of all FLAMINGO galaxies. The red color curves display the fit to the observed star-forming main sequence (SFMS) of galaxies given in \cite{Popesso.etal.2023}. Here, we have shown star formation rates and stellar masses within spherical apertures of 30~pkpc, and flux densities using H13 model. At all redshifts, SMGs tend to fall on or above the SFMS of galaxies. Most of the $z \le 1$ or $z > 6$ SMGs are starburst galaxies.}
    \label{fig:sfr_stellar_mass}
\end{figure*}

SMGs are characterized by their star formation rates and dust mass content. For example, a highly star-forming galaxy with negligible dust content will produce a smaller (sub)mm flux than a galaxy with a low star formation rate but a large dust content. These effects can move SMGs on the star formation rate and stellar mass plane with the evolution of dust in the Universe. To investigate the evolution of SMGs on the star formation rate $-$ stellar mass plane, in Fig.~\ref{fig:sfr_stellar_mass}, we plot the median star formation rate and stellar mass relation of SMGs in the FLAMINGO simulation from $z=6$ to $z=0.5$. For a detailed study, we have grouped SMGs into four flux density ranges shown with yellow ($\rm 1 \le S_{850}/mJy < 2$), orange ($\rm 2 \le S_{850}/mJy < 3$), green ($\rm 3 \le S_{850}/mJy < 5$), and pink ($\rm S_{850} > 5~mJy$) color curves. The respective shaded regions around the median relations represent the 16th to 84th percentile range. The gray color maps in the background show the distribution of all the galaxies in FLAMINGO. Note that we have shown here the star formation rates and stellar masses within a spherical aperture of 30~pkpc radius, which is used for submillimeter flux density calculation in this work. Given the remarkable consistency of the H13 predictions with observations of the SMG population, we only show flux densities obtained from the H13 relation. For comparison, we over-plot observed fitting relations for the star-forming main sequence (SFMS) of all galaxies from \cite{Popesso.etal.2023} in red color.

First, Fig.~\ref{fig:sfr_stellar_mass} shows that the FLAMINGO reproduces the SFMS from \cite{Popesso.etal.2023}, between $z=5.0-0.5$ with slight deviations at redshifts $z > 5$.
If we now focus on the SMG population, we find that almost all the SMGs brighter than 3 mJy lie above the SFMS throughout the cosmic evolution. This suggests that the bright SMGs are always starburst galaxies
On the other hand, majority of SMGs with $\rm 1 \le S_{850}/mJy < 3$ lie on the SFMS in the redshift interval $z=5.0-1.5$, whereas they lie above SFMS for $z \le 1$ or $z > 6$, belonging to starburst galaxies. Our $z \le 1$ redshift results are consistent with \cite{Michalowski.etal.2017} who reported that all the $z < 1$ SCUBA2 SMGs above the SFMS.
Moreover, the four groups of flux densities in Fig.~\ref{fig:sfr_stellar_mass} remain significantly distinct from $z=5.0$ to $z=0.5$, specifically for massive galaxies ($\rm M_{*} > 2 \times 10^{11}~\Msun$). This indicates a strong correlation between (sub)mm flux density and star formation rate. However, this correlation weakens towards low-mass SMGs ($\rm M_{*} < 2 \times 10^{11}~\Msun$) and for $z > 5.0$ SMGs, where we can notice significant overlap among the four groups. Additionally, all the SMGs (indeed all galaxies) in FLAMINGO at $z > 6.0$ have stellar masses $\rm < 2 \times 10^{11}~\Msun$ (for brevity, $z > 6.0$ redshifts are not shown here).

At a given stellar mass, the median star formation rates of the four SMG groups described in Fig.~\ref{fig:sfr_stellar_mass} slowly decrease with decreasing redshift. The high redshift SMGs show higher star formation rates than the low redshift SMGs for the same flux density. This is the effect of dust enrichment in the Universe, which counter-balances the flux density reduction due to the evolution of the star formation rate. \cite{Magliocchetti.etal.2013, Wilkinson.etal.2017} report downsizing of halos for a given star formation rate. From their analysis, they inferred that the host halos of SMGs at lower redshift have a lower mass than those at higher redshift for the same star formation activity. Our analysis does not appear to show such a downsizing effect. We found SMGs with a range of stellar masses for a given star formation rate, specifically SMGs with $\rm 1 \le S_{850}/mJy < 3$.

\subsection{Predictions for flux density functions}
\label{sec:flux_density_func}

\begin{figure}
    \centering
    \includegraphics[width=\columnwidth]{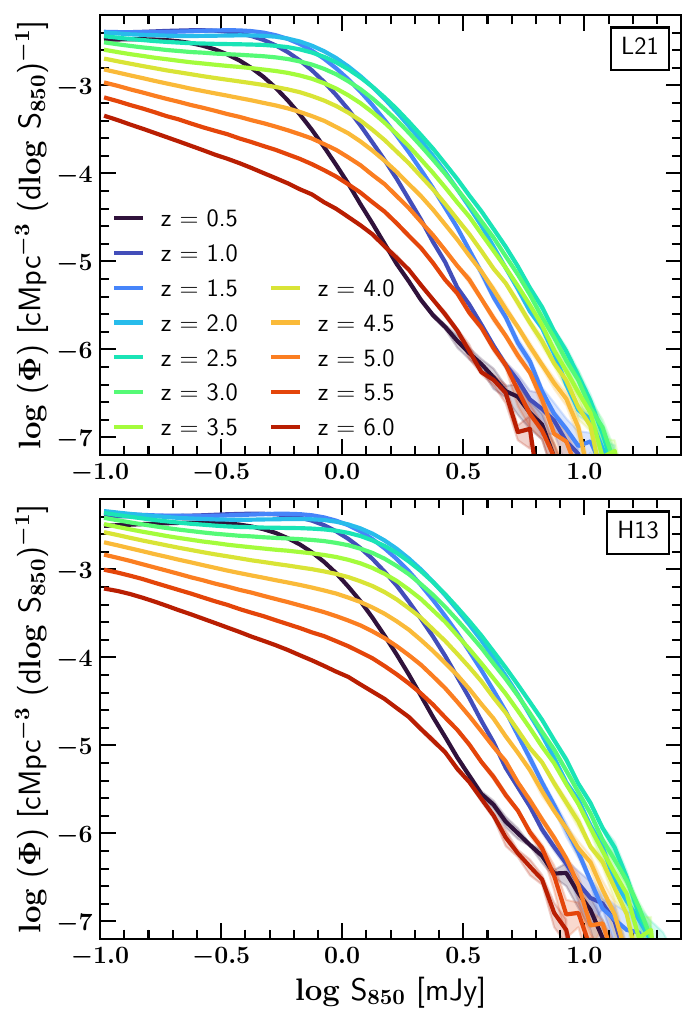}
    \caption{Flux density functions of submm galaxies in the FLAMINGO simulation for $z = 0.5 - 6.0$ redshift range. Top panel shows prediction using L21 relation and bottom panel displays predictions assuming H13 relation. The brighter sources increase at higher rate than the fainter sources from $z=6.0$ to $z=2.5$ followed by a rapid depletion of bright sources for $z<2.5$.}
    \label{fig:flux_dens_func}
\end{figure}

In this section, we use the submm galaxies population in FLAMINGO simulation to predict their flux density functions at different redshifts. Fig.~\ref{fig:flux_dens_func} shows the flux density functions from redshift $z=0.5$ to 6.0 using the L21 model in the top panel and the H13 model in the bottom panel. 
At all flux densities shown here, the submm population grows monotonically from $z=6.0$ to $z=2.5$ as is evident in the two panels. The number of brighter sources increases at higher rate than the fainter sources. For example, from redshift $z=6.0$ to $z=2.5$, the submm population grows by about a factor of $\sim 10$ at 0.1 mJy, and about a factor of $\sim 60$ at 1 mJy. After the that bright submm population depletes rapidly with decreasing redshift. This evolution of submm population changes the shape of flux density function across cosmic time. The low-redshift flux density functions show prominent knee as compare to high-redshift flux density functions. Also, the location of the knee moves towards lower flux density side with decreasing redshift as clearly visible in two panels.

The increasing submm population from $z=6.0$ to $z=2.5$ in both the panels is expected, given the dependency of modeled submm flux on the star formation rate, which is strongly linked with the cosmic SFRD. For a quantitative characterization of this evolution, we fit the Schechter function defined as,
\begin{equation}
    \phi(S)~d\log S = ln(10)~\phi^{*}~\left(\frac{S}{S^{*}}\right)^{\alpha+1}~\exp{\left(-\frac{S}{S^{*}} \right)}~d\log S,
    \label{eqn:schechter_func}
\end{equation}
where $\phi^{*}$ is the normalization parameter, S$^{*}$ is the characteristic flux density, and $\alpha$ is the  slope parameter. 

\begin{figure}
    \centering
    \includegraphics[width=\columnwidth]{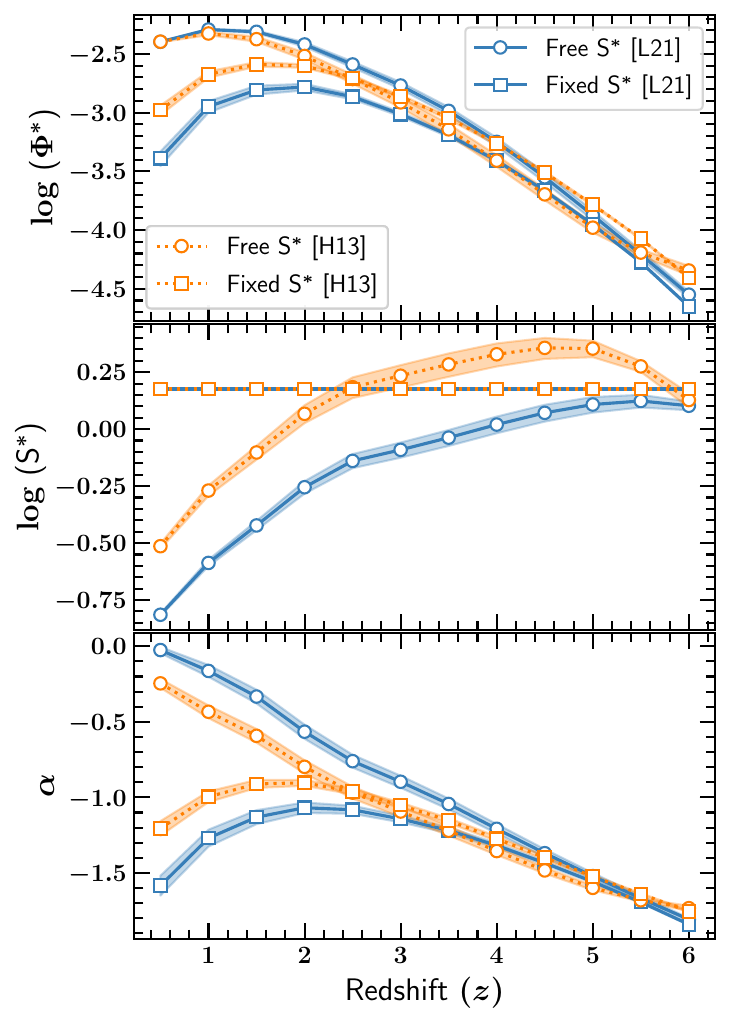}
    \caption{Redshift evolution of the Schechter function fitting parameters for the flux density function shown in Fig.~\ref{fig:flux_dens_func}. The solid blue and dotted orange curves with circles represent, respectively, the fitting parameters for the L21 and H13 models. To exhibit the evolution of the flux density functions, we also show the fitting parameters for an arbitrarily fixed value of S$^{*}=1.5$ mJy using solid blue and dotted orange curves with boxes for the L21 and H13, respectively. The shaded regions around the curves represent fitting uncertainties.}
    \label{fig:flux_dens_func_fit}
\end{figure}

In Fig.~\ref{fig:flux_dens_func_fit}, we show the redshift evolution of the Schechter function fitting parameters $\phi^{*}$ (top panel), S$^{*}$ (middle panel), and $\alpha$ (bottom panel) for the flux density functions. The blue and orange lines with circles correspond to the fitting parameters for the L21 and H13 models, respectively, when the three parameters are left unconstrained. The lines with boxes show results when we arbitrarily fix S$^{*}=1.5$ mJy. 
We constrain the fits in the range of $S_{850}=[0.1-25]$~ mJy for the H13 and $S_{850}=[0.1-25]/1.75$~mJy for the L21 (since the H13 model predicts about a factor of 1.75 higher flux densities than the L21 model on average).

Qualitatively, the parameters of the Schechter function demonstrate comparable trends for both the L21 and H13 relations. We find that the normalization parameter $\phi^{*}$ increases with decreasing redshift when all parameters are free during fitting. However, for a fixed value of S$^{*}$, the normalization parameter increases until $z \approx 2$ and then decreases indicating the growth of the SMG population, as evident from the flux density functions shown in Fig.~\ref{fig:flux_dens_func}. When we leave the characteristic flux density S$^{*}$ as a free parameter, it changes very slowly between $z\simeq 6.0-2.5$ and then decreases sharply with decreasing redshift for $z<2.5$, an indication of the reduction in the submm-bright population. A similar qualitative evolution has been reported for the infrared luminosity functions of galaxies \citep[e.g.,][]{Gruppioni.etal.2013, Fujimoto.etal.2023, Traina.etal.2024}. 
We find that the slope parameter $\alpha$ increases linearly with decreasing redshift for free S$^{*}$, an indication of flattening towards faint-end flux density function as seen in Fig.~\ref{fig:flux_dens_func}. For a fixed S$^{*}$, it increases until $z=2.0$ and then decreases with decreasing redshift.

\section{Predictions for the TolTEC UDS and LSS surveys}
\label{sec:toltec_prediction}

\begin{figure*}
    \centering
    \includegraphics[width=\textwidth]{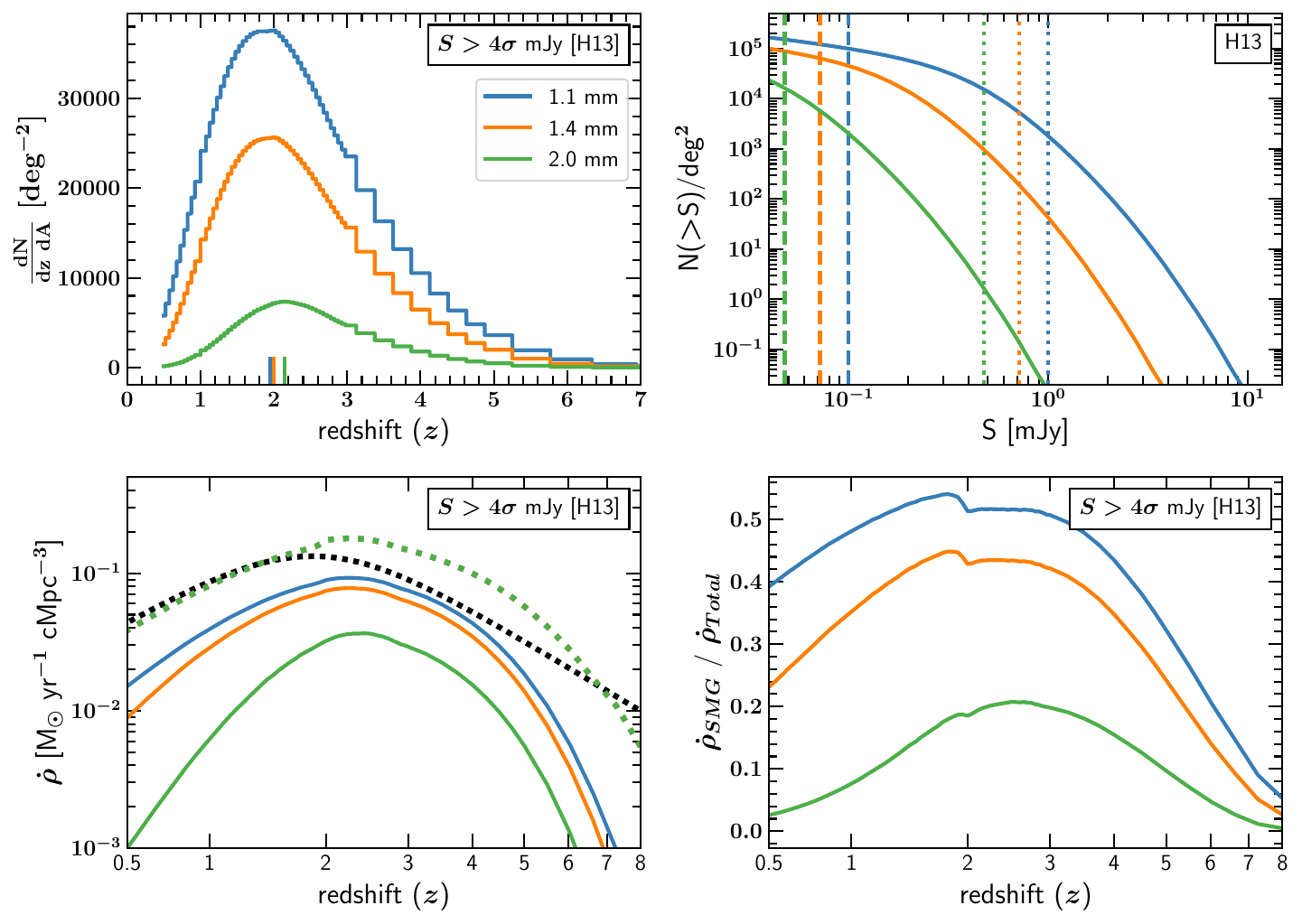}
    \caption{TolTEC UDS survey predictions for redshift distributions (top-left panel), source number counts (top-right panel), SFRDs (bottom-left panel), and ratio of submm SFRDs to the total cosmic SFRDs (bottom-right panel) using the H13 relation and the FLAMINGO simulation. The blue, orange and green curves respectively represent $1.1~\mm$, $1.4~\mm$, and $2.0~\mm$ wavelength bands. Vertical dashed and dotted lines in top-right panel indicate the $4\sigma$ depths of the TolTEC UDS and LSS surveys, respectively. In the bottom-left panel, the green dotted curve is the total cosmic SFRD and solid curves are (sub)mm SFRDs of FLAMINGO, and the black dotted curve is the \cite{Madau.Dickinson.2014} fit to observations. The TolTEC $\rm 0.8~deg^{2}$ UDS survey at $1.1~\mm$ is expected to observe $\approx 80,000$ sources above the $4\sigma$ detection limit, which will contribute to about 50\% of the cosmic SFRD at $z\approx2.5$.} 
    \label{fig:toltec_4_H}
\end{figure*}

The remarkable good agreement between the predictions of the FLAMINGO simulation and observations of (sub)mm bright galaxies, particularly using the H13 relation, allows us to make predictions for upcoming observational surveys. 
In particular, the future TolTEC\footnote{\url{http://toltec.astro.umass.edu/science\_legacy\_surveys.php}} camera \citep{Wilson.etal.2020} at the 50-m Large Millimeter Telescope \citep[LMT: ][]{Hughes.etal.2020} will provide new insights into dust-obscured star-forming galaxies. 
Its mapping speeds greater than 2~deg$^2$/mJy$^2$/hr and its high angular resolution will allow conducting two public legacy surveys at wavelengths of $1.1~\mm$, $1.4~\mm$, and $2.0~\mm$: the $\rm 0.8~deg^{2}$ Ultra Deep Survey (UDS), and the $\rm 40-60~deg^{2}$ Large Scale Survey (LSS). 

The UDS will reach an optimal depth of $\sigma_{1.1} \approx 0.025$ mJy at $1.1~\mm$, $\sigma_{1.4} \approx 0.018$ mJy at $1.4~\mm$, and $\sigma_{2.0} \approx 0.012$ mJy at $2.0~\mm$, whereas the depth of LSS will be 10 times shallower than UDS. Here we use our FLAMINGO submm galaxies catalogue to make predictions for the TolTEC UDS and LSS surveys. Although we provide preliminary predictions, we plan to expand our forecasts for the UDS and LSS surveys in a future study. For this analysis, we scale the flux densities at $850~\mum$ to the TolTEC bands using equation~(\ref{eqn:flux_conversion}).

Fig~\ref{fig:toltec_4_H} shows the predicted redshift distributions, source number counts, SFRD of submm galaxies, and ratio of submm SFRD to cosmic SFRD for the TolTEC UDS survey above the $4\sigma$ detection threshold. For brevity, we only show the predictions assuming the H13 relation, as H13 reproduces the $850~\mum$ observations quite well. The blue, orange, and green color curves represent, respectively, $1.1~\mm$, $1.4~\mm$, and $2.0~\mm$ predictions. The vertical dashed lines in the top-right panel indicate the $4\sigma$ limits of the UDS survey, and the vertical dotted lines indicate the $4\sigma$ limits of the LSS survey. The yellow dashed curve in the bottom-left panel shows the total cosmic SFRD in the FLAMINGO simulation, and the black dashed curve represents \cite{Madau.Dickinson.2014} fit to the observations. 

From the top-left panel of Fig~\ref{fig:toltec_4_H} we see that the median values of the redshift distributions in the three TolTEC bands are between $z\approx 1.95-2.15$. Due to survey depth limits, we expect that the $1.1~\mm$ distribution will have a slightly lower median value than the $2.0~\mm$ distribution because brighter (sub)mm galaxies tend to lie at higher redshifts (see Section~\ref{sec:redshift_distribution}). Studies suggest that $2~\mm$ observations preferentially select high redshift galaxies \cite{Casey.etal.2021}. However, our analysis shows only marginal difference between median redshifts of $1.1~\mm$ and $2.0~\mm$ distributions likely because we have modeled them by scaling $850~\mum$ fluxes. The top-right panel of Fig.~\ref{fig:toltec_4_H} shows the FLAMINGO predictions that the TolTEC UDS survey will observe about $13,000$ galaxies at $2.0~\mm$, $51,200$ galaxies at $1.4~\mm$, and $80,000$ galaxies at $1.1~\mm$. Similarly, we expect that the TolTEC LSS survey will observe about $60-90$ galaxies at $2.0~\mm$, $7,500-11,300$ galaxies at $1.4~\mm$, $73,000-109,400$ galaxies at $1.1~\mm$. 
The bottom-left panel of Fig.~\ref{fig:toltec_4_H} compares (sub)mm SFRDs for the three TolTEC bands with the total cosmic SFRD. The bottom-right panel shows their relative contributions. At $z\approx 2.5$, we expect that the TolTEC UDS survey will reveal about $20\%$ of the total SFRD in the $2.0~\mm$ band, $40\%$ SFRD in the $1.4~\mm$ band, and about $50\%$ SFRD in the $1.1~\mm$ band. Note that the contribution of submm galaxies to cosmic SFRD in three bands is not mutually exclusive because all the galaxies observed in longer wavelength band will also be observed in shorter wavelength band. Our predictions can slightly be affected by the cosmic variance and multiple source blending.

\section{Conclusions}
\label{sec:conclusions}
In this work, we utilize  three state-of-the-art cosmological hydrodynamical simulations, EAGLE, IllustrisTNG and FLAMINGO, to model (sub)mm flux densities in galaxies, compare with observations, and make predictions for the upcoming TolTEC surveys. The main results are summarized as follows.

\begin{enumerate}
\item We take advantage of the EAGLE RT calculations from \cite{Camps.etal.2018} to test the H13 \citep{Hayward.etal.2013}, L21 \citep{Lovell.etal.2021}, and C23 \citep{Cochrane.etal.2023} parametric models for the SCUBA2\_850 flux density measurement using the SFR and dust mass of galaxies.
Furthermore, we test 
the predictions of these models on the observational sample of 892 SMGs compiled from the literature, where SFRs and dust masses are available.
As shown in Table~\ref{tab:s850_parameters}, the main difference between the parametric models is the normalization factor, which impacts the number of galaxies exceeding the flux density S$_{850} > $1~mJy. Our analysis shows that L21 and C23 models exhibit remarkable consistency with the EAGLE RT calculations whereas the H13 model over-estimates the flux densities when compared to EAGLE RT results (see Fig.~\ref{fig:RT_Hay13_Lov21_comparison}).
In contrast, the H13 model reproduces the observed flux densities much better than L21 and C23, which tend to under-predict them (see Fig.~\ref{fig:models_vs_observations}). 
Based on these findings we choose to proceed our study with H13 and L21 parametric models.



\item We find that the FLAMINGO simulation reproduces the shape and median of the observed redshift distribution of SMGs for  H13 and L21 parametric models. 
On the other hand, the redshift distribution of SMGs in IllustrisTNG peaks at a higher redshift than observed for both parametric models (see Fig.~\ref{fig:redshift_dist_all_sim}). 
Interestingly, our results show that using the H13 model, the FLAMINGO simulation can reproduce the source number counts of SMGs from observations (see Fig.~\ref{fig:number_counts_all_sim}). This is not the case for the IllustrisTNG and EAGLE simulations which underpredict the trend. 


\item We find that the contribution of submm-bright galaxies (S$_{850} >1$ mJy) to the cosmic SFRD reaches about 27\% at $z=2.6$ in 
the FLAMINGO simulation. This finding is consistent with observational estimates \citep[see,][and our Fig.~\ref{fig:sfr_dens_all_sim}]{Swinbank.etal.2014, Dudzeviciute.etal.2020} . 

\item All the SMGs with $\rm S_{850} > 3~mJy$ are starburst galaxies and lie above the SFMS of galaxies throughout cosmic evolution. SMGs with $\rm 1 \le S_{850}/mJy < 3$ fall at the SFMS in the redshift interval $z=5.0-1.5$ and above it for $z \le 1$ or $z \geq 6.0$ (see Fig.~\ref{fig:sfr_stellar_mass}). At all redshifts, the flux densities of high-mass SMGs ($M_{*} > 2 \times 10^{11}~\Msun$) show a strong correlation with their star formation rates, while this correlation weakens for low-mass SMGs. In contrast to \cite{Magliocchetti.etal.2013, Wilkinson.etal.2017}, for the same star formation rate, we do not find a downsizing effect in the FLAMINGO SMGs.

\item We use FLAMINGO to make predictions for the flux density function with redshift. Our results reveal a rise in SMG abundance from $z=6$ to $z=2.5$ (see Fig.~\ref{fig:flux_dens_func}). The Schechter function fits the flux density functions for a fixed characteristic flux density ($S^{*}$) showing a linearly increasing slope parameter ($\alpha$) from $z=6$ to $z=2.5$ and then decreasing. Between $z=2.5-0$, we find that the depletion of SMGs accelerates with increasing flux density indicating rapid fading of the brightest submm phase (see Fig.~\ref{fig:flux_dens_func_fit}).

\item We provide predictions for the upcoming TolTEC $\rm 0.8~deg^{2}$ Ultra Deep Surve (UDS) and $\rm 40-60~deg^{2}$ Large Scale Survey (LSS) at the $4\sigma$ detection threshold. Our results suggest that the UDS survey will observe about $13,000$ galaxies at $2.0~\mm$, $51,200$ galaxies at $1.4~\mm$, and $80,000$ galaxies at $1.1~\mm$. The UDS survey will observe a large fraction of low-mass SMGs. About 13,000 bright galaxies will be observed in three bands providing a large sample for studying dust properties. On the other hand, the LSS survey will observe about $60-90$ galaxies at $2.0~\mm$, $7,500-11,300$ galaxies at $1.4~\mm$, $73,000-109,400$ galaxies at $1.1~\mm$. Furthermore, the UDS survey will reveal (sub)mm galaxies that contribute about 50\% to the cosmic SFRD at $z\approx 2.5$ in $1.1~\mm$ band (see Fig~\ref{fig:toltec_4_H}).


\end{enumerate}
\begin{acknowledgements}
AK and MCA acknowledge support from ALMA fund with code 31220021 and from ANID BASAL project FB210003. ADMD thanks Fondecyt for financial support through the Fondecyt Regular 2021 grant 1210612. We thank CC Hayward, SM Stach, E van Kampen, P Camps, and M Baes for  comments and suggestions. HSH acknowledges the support of the National Research Foundation of Korea (NRF) grant funded by the Korea government (MSIT), NRF-2021R1A2C1094577, Samsung Electronic Co., Ltd. (Project Number IO220811-01945-01), and Hyunsong Educational \& Cultural Foundation. J.L. is supported by the National Research Foundation of Korea (NRF-2021R1C1C2011626).
We make use of Matplotlib \citep{matplotlib2007}, Numpy \citep{numpy2020}, Pandas \citep{pandas2010, pandas2020}, Scipy \citep{Scipy2020}, WebPlotDigitizer in this work. 
This work used the RAGNAR computing facility available at Universidad Andr\'es Bello.
This work used the DiRAC@Durham facility managed by the Institute for Computational Cosmology on behalf of the STFC DiRAC HPC Facility (\url{www.dirac.ac.uk}). The equipment was funded by BEIS capital funding via STFC capital grants ST/P002293/1, ST/R002371/1 and ST/S002502/1, Durham University and STFC operations grant 
ST/R000832/1. DiRAC is part of the National e-Infrastructure.
We acknowledge the Virgo Consortium for making their simulation data available. The EAGLE simulations were performed using the DiRAC-2 facility at Durham, managed by the ICC, and the PRACE facility Curie based in France at TGCC, CEA, Bruy\`eres-le-Ch\^atel.
The IllustrisTNG simulations were undertaken with compute time awarded by the Gauss Centre for Supercomputing (GCS) under GCS Large-Scale Projects GCS-ILLU and GCS-DWAR on the GCS share of the supercomputer Hazel Hen at the High Performance Computing Center Stuttgart (HLRS), as well as on the machines of the Max Planck Computing and Data Facility (MPCDF) in Garching, Germany.
\end{acknowledgements}
\bibliographystyle{aa}
\bibliography{main_aanda.bib}
\begin{appendix}

\section{The effect of box size}
\label{appendix:res_mass_effect}
\begin{figure*}
    \centering
    \includegraphics[width=\textwidth]{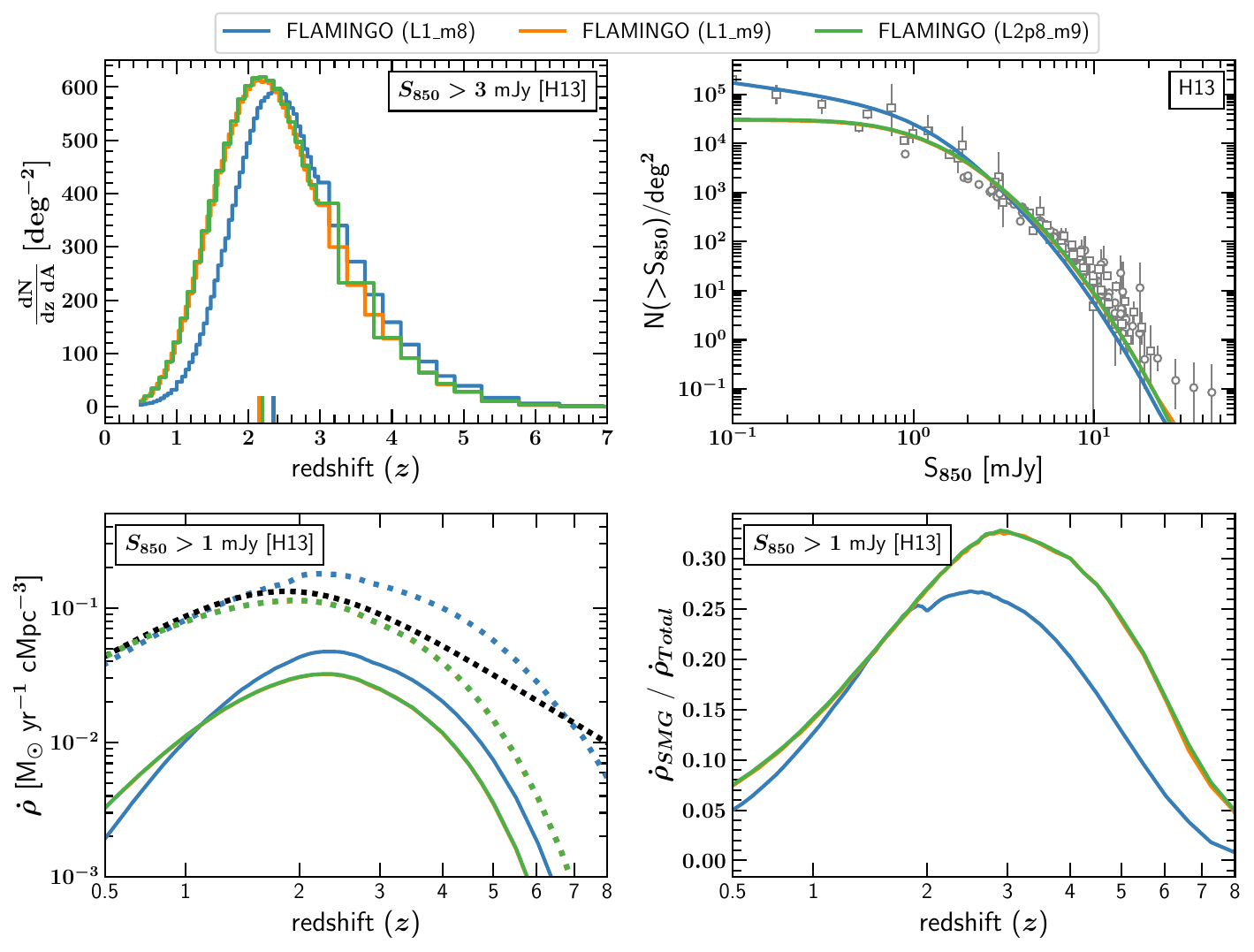}
    \caption{Comparison of the redshift distributions, source number counts, SFRDs, and ratio of submm-bright SFRDs to cosmic SFRDs using the H13 relation for the three fiducial FLAMINGO simulations having different resolution and box size. Simulations L1\_m8, L1\_m9, and L2p8\_m9 are respectively shown with blue, orange and green colors. The gray symbols in the top-right panel are observational data. The small bars at the bottom of top-left panel indicate the medians of the respective distributions. In the bottom-left panel, dotted curves are the cosmic SFRDs, solid curves are submm-bright SFRDs, and the black dotted curve is the \cite{Madau.Dickinson.2014} fit to observations. The remarkable overlap between L1\_m9 and L2p8\_m9 shows that a 1~cGpc box side length is large enough for modeling the (sub)mm galaxy population. The top-right panel shows that higher resolution improves fainter end source number counts.}
    \label{appfig:res_box_H}
\end{figure*}

This work particularly focuses on one of the FLAMINGO flagship simulations in a box with side length 1~cGpc (L1\_m8). In this appendix, we investigate the effect of box size on our findings. For this purpose, we utilised two other FLAMINGO simulations: one in the same box size of 1~cGpc (L1\_m9), and another in a larger box size of 2.8~cGpc (L2p8\_m9). Both of these simulations have the same mass resolution, but have 8 times lower mass-resolution than the L1\_m8 simulation. In Fig.~\ref{appfig:res_box_H}, we compare the redshift distributions, source number counts, SFRDs, and ratio of submm-bright SFRDs to their cosmic SFRDs assuming the H13 model. Simulations L1\_m8, L1\_m9, and L2p8\_m9 are respectively shown with blue, orange, and green colors. To visualize the importance of resolution on the source number counts towards the fainter end, we have over-plotted the observational data.

All the panels of Fig.~\ref{appfig:res_box_H} show remarkable consistency with increasing box size (the orange and green color curves are barely distinguishable). This confirms that the 1~cGpc side length box is sufficiently large to model the (sub)mm galaxy population in cosmological simulations. Further, increasing the resolution improves the source number counts towards the faint end while having only a marginal effect on the bright end, as is evident from the blue and orange curves in top-right panel. There is some effect of resolution on the submm SRFD and cosmic SFRD, and thus on the redshift distribution. This is a consequence of the self-consistent evolution of FLAMINGO simulations which were not calibrated to the cosmic SFRD. The simulation with increased resolution (L1\_m8) shows a higher cosmic SFRD for $z > 2$. However, this increase in the cosmic SFRD does not significantly increase the number of submm-bright galaxies, as is evident from the redshift distribution in the top-left panel. This effect results in reducing the contribution of submm galaxies to the cosmic SRFD compared to the other FLAMINGO simulations.

\section{The effect of the galaxy formation prescription}
\label{appendix:gal_form_pres_effect}
\begin{figure*}
    \centering
    \includegraphics[width=\textwidth]{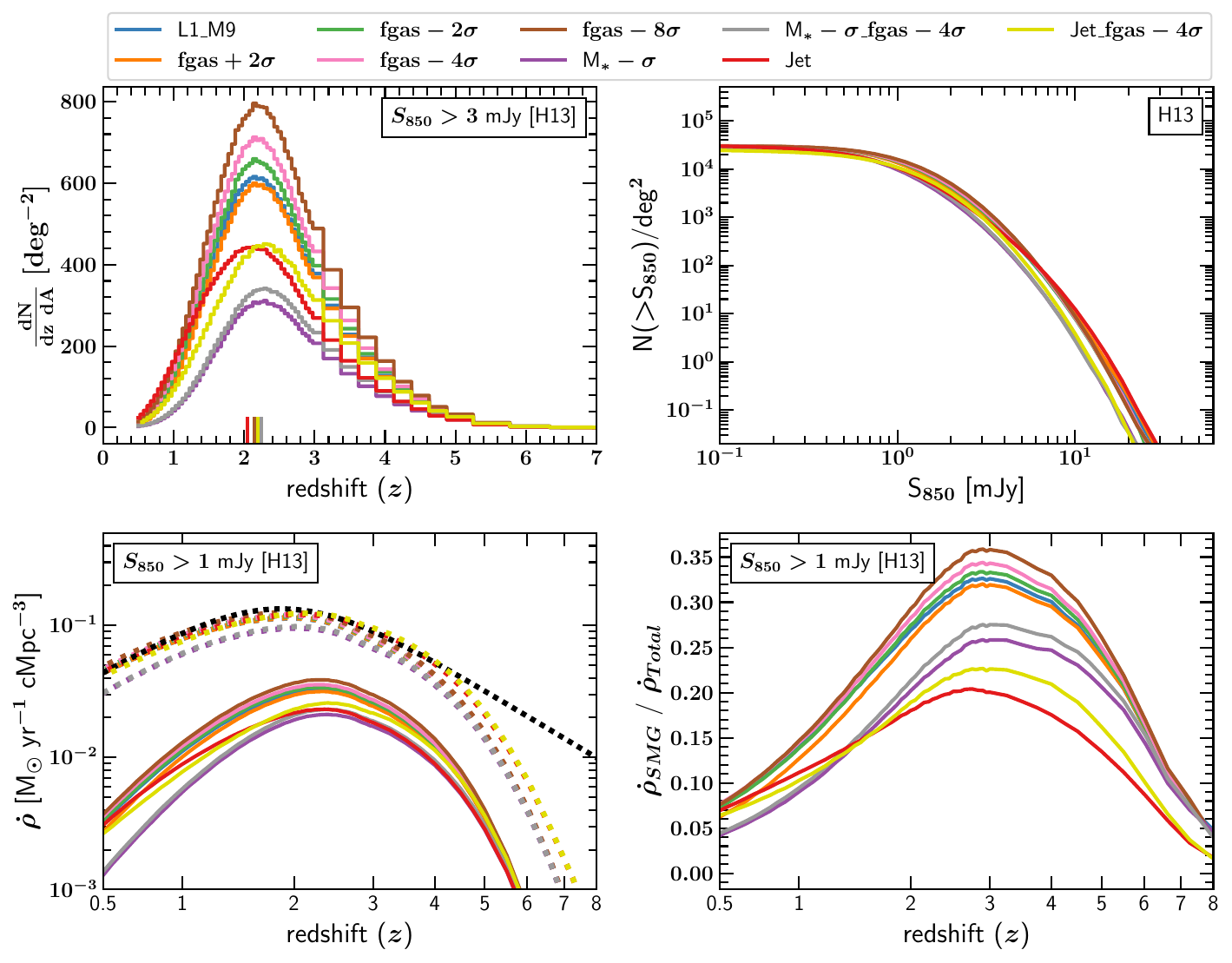}
    \caption{As Fig.~\ref{appfig:res_box_H} but comparing FLAMINGO simulations using different galaxy formation prescriptions. Galaxy formation prescription with strongest AGN predicts excess submm-bright sources contributing higher to the cosmic SFRD than other galaxy formation prescriptions.}
    \label{appfig:GFP_H}
\end{figure*}

To account for the systematic uncertainties in the observed stellar mass and cluster gas fractions, FLAMINGO includes eight variations of the galaxy formation prescriptions. These simulations with varying galaxy formation prescriptions are performed in a 1~cGpc box \citep[for details see][]{Schaye.etal.2023}. In this appendix, we make use of these simulations to test the effect of the galaxy formation prescription on the submm galaxy population. Fig~\ref{appfig:GFP_H} shows the comparison of redshift distributions, source number counts, SFRDs, and ratio of submm SFRDs to cosmic SFRDs assuming the H13 relation for the simulations with varying galaxy formation prescriptions. In Table~\ref{apptab:gal_form_pres}, we list the names of the simulations and calibration parameters which they are calibrated for. 
Note that AGN feedback particularly affects the cluster gas fraction, and SN feedback particularly affects galaxy stellar mass.

\begin{table}
    \centering
    \caption{Name of the simulations with varying galaxy formation prescriptions and the observed parameters used for calibration. The parameter f$_{gas}$ indicates the shift with respect to the observed gas fraction in clusters, M$_{*}$ indicates the shift with respect to the observed stellar mass of galaxies, and $\sigma$ are the corresponding systematic uncertainties in the data.}
    \begin{tabular}{|l|l|}
    \hline
    Identifier in \cite{Schaye.etal.2023} & Calibrated to \\
    \hline
    L1\_M9 & f$_{gas}$, M$_{*}$ \\
    $\rm fgas+2\sigma$ & f$_{gas}+2\sigma$, M$_{*}$ \\
    $\rm fgas-2\sigma$ & f$_{gas}-2\sigma$, M$_{*}$ \\
    $\rm fgas-4\sigma$ & f$_{gas}-4\sigma$, M$_{*}$ \\
    $\rm fgas-8\sigma$ & f$_{gas}-8\sigma$, M$_{*}$ \\
    M$_{*}-\sigma$ & f$_{gas}$, M$_{*}-\sigma$ \\
    M$_{*}-\sigma$\_$\rm fgas-4\sigma$ & f$_{gas}-4\sigma$, M$_{*}-\sigma$ \\
    Jet & f$_{gas}$, M$_{*}$\\
    Jet\_$\rm fgas-4\sigma$ & f$_{gas}-4\sigma$, M$_{*}$\\
    \hline
    \end{tabular}
    \label{apptab:gal_form_pres}
\end{table}

The top-left panel of Fig~\ref{appfig:GFP_H} shows that the simulation with the lowest cluster gas fractions, which are mainly controlled by the AGN feedback, predicts the highest number of submm sources brighter than 3 mJy. The low galaxy mass function models predict lower numbers of submm-bright sources. Additionally, the cosmic SFRDs of low galaxy mass function models are lower, as is evident from the bottom-left panel. The ratio of submm-bright SFRDs to cosmic SFRDs in the bottom-right panel clearly indicates that for the models using jet-like AGN feedback, submm galaxies contribute the least to the cosmic SRFD. The predictions of low galaxy mass function models fall in between those of the low gas fraction models. On the other hand, the source number counts in top-right panel are marginally affected by the choice of galaxy formation prescription.

\section{The effect of cosmology}
\label{appendix:gal_form_cosmo_effect}
\begin{figure*}
    \centering
    \includegraphics[width=\textwidth]{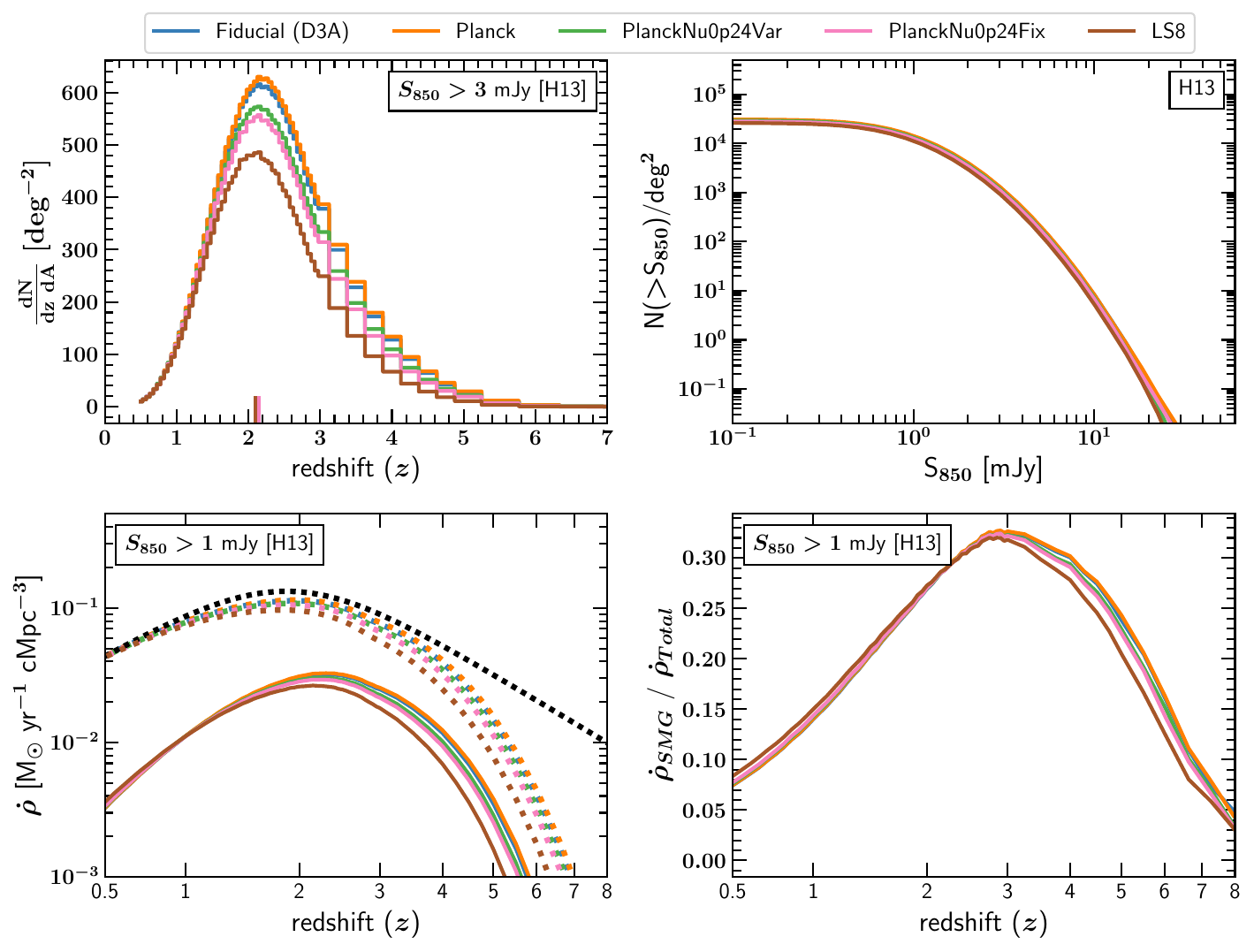}
    \caption{As Fig.~\ref{appfig:res_box_H} but comparing FLAMINGO simulations assuming different cosmological parameters. Predictions for all the cosmologies are quite consistent given their cosmic SFRDs.}
    \label{appfig:GFC_H}
\end{figure*}

Along with the fiducial cosmology, the four variations of cosmology in the FLAMINGO suite of simulations provide us with the opportunity to test the effect of cosmological parameters on the predicted submm galaxies. Similar to the variations of galaxy formation prescriptions, the variations of cosmologies are also simulated in 1~cGpc boxes. Fig.~\ref{appfig:GFC_H} illustrates the effect of variation in cosmology on the redshift distributions, source number counts, SFRDs, and ratio of submm SFRDs to cosmic SFRDs assuming the H13 relation. For the cosmological parameters, we refer the reader to Table 4 in \cite{Schaye.etal.2023}. Briefly, Dark Energy Survey parameters \citep{Abbott.etal.2022} are \textit{Fiducial} cosmology (\textit{D3A}), Planck collaboration parameters \citep{Planck_Collaboration.2020} with minimum allowed neutrino mass are \textit{Planck} cosmology, adjusted Planck collaboration parameters with maximum allowed neutrino mass is \textit{PlanckNu0p24Var} cosmology, fixed Planck collaboration parameters with maximum allowed neutrino mass is \textit{PlanckNu0p24Fix} cosmology, and Lensing cosmology parameters \citep{Amon.etal.2023} are \textit{LS8} cosmology.

All the panels of Fig.~\ref{appfig:GFC_H} display consistent predictions at low redshift, irrespective of the cosmology. Furthermore, the \textit{D3A} and \textit{Planck} cosmologies show similar predictions at all redshifts. The other cosmologies, which have lower values of $\sigma_{8}$, predict reduced numbers of submm-bright galaxies towards higher redshift, which is consistent with their relatively lower cosmic SFRDs. The ratio of the submm-bright SFRD to cosmic SFRD in the bottom-right panel exhibits very similar contributions of submm-bright galaxies to the cosmic SFRD for all the cosmologies.

\end{appendix}
\end{document}